\documentclass[acmtog]{acmart}

\AtBeginDocument{%
  \providecommand\BibTeX{{%
    \normalfont B\kern-0.5em{\scshape i\kern-0.25em b}\kern-0.8em\TeX}}}

\newcommand{\secref}[1]{\S~\ref{sec:#1}}
\newcommand{\tabref}[1]{Tab.~\ref{tab:#1}}
\newcommand{\figref}[1]{Fig.~\ref{fig:#1}}
\renewcommand{\eqref}[1]{Eq.~(\ref{eq:#1})}

\newcommand{\etal}{et al.~}

\newcommand{\mbx}{\ensuremath{\mathbf{x}}}

\newcommand{\mbz}{\ensuremath{\mathbf{z}}}
\newcommand{\mbv}{\ensuremath{\mathbf{v}}}

\setcopyright{acmcopyright}
\copyrightyear{xxxx}
\acmYear{xxxx}
\acmDOI{XXXXXXX.XXXXXXX}

\acmJournal{TOG}
\acmVolume{37}
\acmNumber{4}
\acmArticle{111}
\acmMonth{8}
\usepackage{balance}
\usepackage[percent]{overpic}
\usepackage{placeins}
\usepackage{algorithm}
\usepackage{algpseudocode}
\usepackage{amsmath}
\usepackage{multirow}
\usepackage{makecell}
\acmSubmissionID{679}

\begin{document}

\title{Online Photon Guiding with 3D Gaussians for Caustics Rendering}

	\author{Jiawei Huang}
	\affiliation{%
		\institution{Chuzhou University and Void Dimensions}
		\country{China}}
  
	\author{Hajime Tanaka}
	\affiliation{%
		\institution{Tohoku University}
		\country{Japan}}
  
	\author{Taku Komura}
	\affiliation{%
		\institution{The University of Hong Kong}
		\country{Hong Kong, China}
	}
 
	\author{Yoshifumi Kitamura}
	\affiliation{%
		\institution{Tohoku University}
		\country{Japan}
	}

\renewcommand{\shortauthors}{Trovato and Tobin, et al.}

\begin{abstract}
In production rendering systems, caustics are typically rendered via photon mapping and gathering, a process often hindered by insufficient photon density. In this paper, we propose a novel photon guiding method to improve the photon density and overall quality for caustic rendering. The key insight of our approach is the application of 3D Gaussian mixtures, used in conjunction with an adaptive light sampler. This combination effectively guides photon emission in expansive 3D scenes with multiple light sources. By employing a 3D Gaussian mixture for each light source, our method precisely models the distribution of the points of interest for photon emission. To sample diffuse emission from the 3D Gaussian mixture at any observation point, we introduce a novel directional transform of the 3D Gaussian, which ensures accurate photon emission guiding. Furthermore, we propose a scene-geometry-based initialization method to improve the optimization efficiency. Our method integrates a global light cluster tree, which models the contribution distribution of light sources to the image, facilitating effective light source selection. We conduct experiments demonstrating that our approach robustly outperforms existing photon guiding techniques across a variety of scenarios, significantly advancing the quality of caustic rendering.
\end{abstract}

\begin{teaserfigure}
\setlength{\abovecaptionskip}{2pt}
\begin{tabular}{rcccccc}
    & Reference&(a) Ours&(b) BG&(c) H2D&(d) vMF & (e) MCMC\\
    \begin{minipage}{0.36\linewidth}
        \centering
        \begin{overpic}[trim = 320 0 200 0, clip,height=5.1cm]{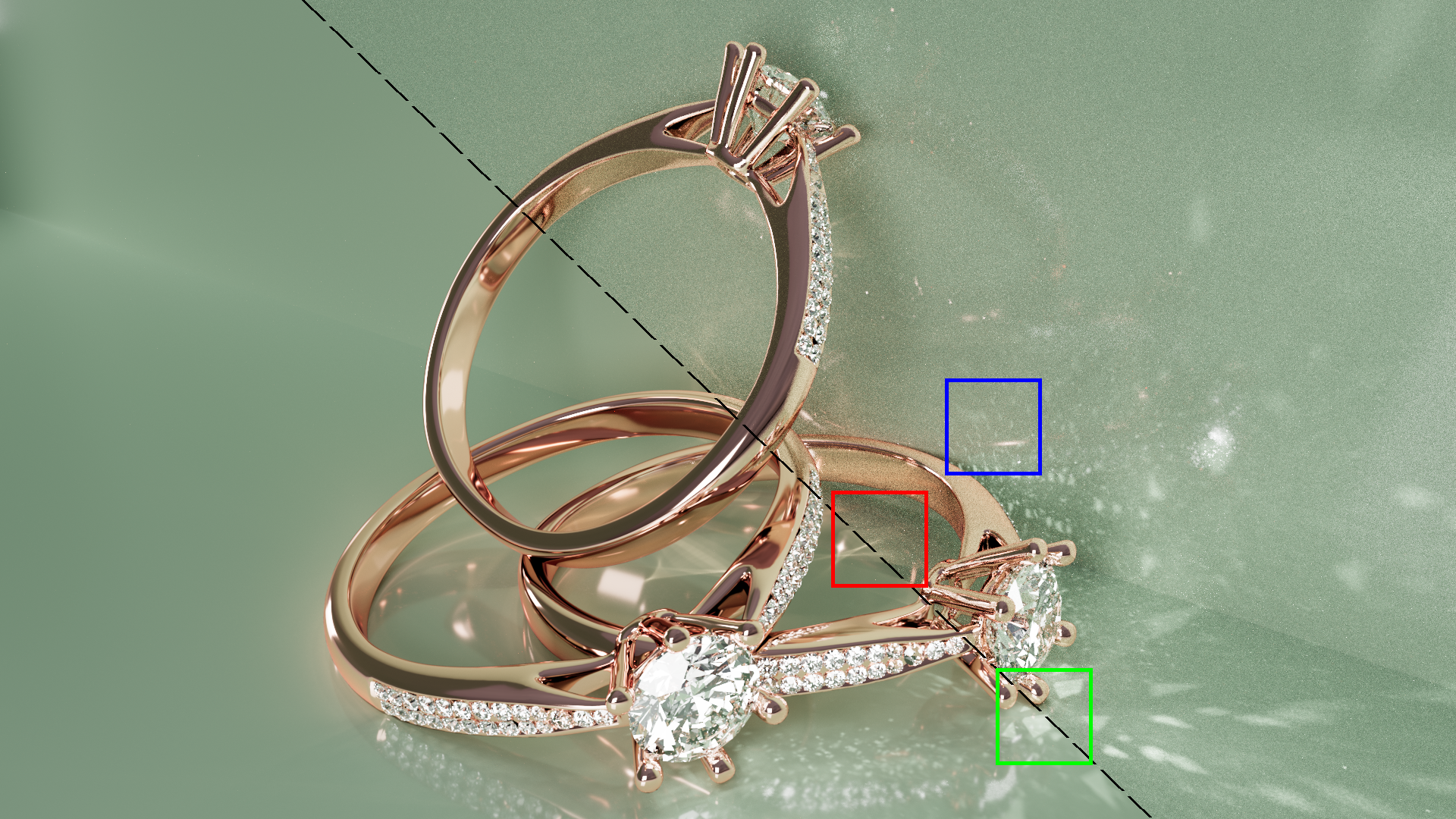}
          \put(1,1){Reference}
          \put(90,72){Ours}
        \end{overpic}
    \end{minipage}
        &
    \begin{minipage}{0.08\linewidth}
        \includegraphics[height=1.7cm]{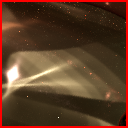}
        \newline
        \includegraphics[height=1.7cm]{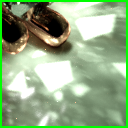}
        \newline
        \includegraphics[height=1.7cm]{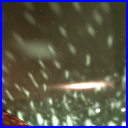}
    \end{minipage}
        &
    \begin{minipage}{0.08\linewidth}
        \includegraphics[height=1.7cm]{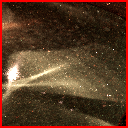}
        \newline
        \includegraphics[height=1.7cm]{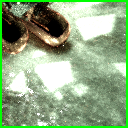}
        \newline
        \includegraphics[height=1.7cm]{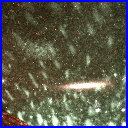}
    \end{minipage}
        &
    \begin{minipage}{0.08\linewidth}
        \includegraphics[height=1.7cm]{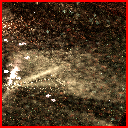}
        \newline
        \includegraphics[height=1.7cm]{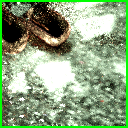}
        \newline
        \includegraphics[height=1.7cm]{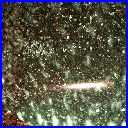}
    \end{minipage}
        &
    \begin{minipage}{0.08\linewidth}
        \includegraphics[height=1.7cm]{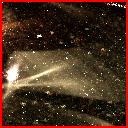}
        \newline
        \includegraphics[height=1.7cm]{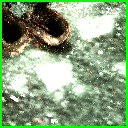}
        \newline
        \includegraphics[height=1.7cm]{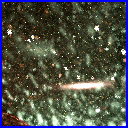}
    \end{minipage}
        &
    \begin{minipage}{0.08\linewidth}
        \includegraphics[height=1.7cm]{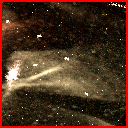}
        \newline
        \includegraphics[height=1.7cm]{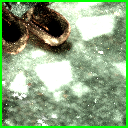}
        \newline
        \includegraphics[height=1.7cm]{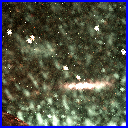}
    \end{minipage}
        &
    \begin{minipage}{0.08\linewidth}
        \includegraphics[height=1.7cm]{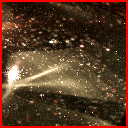}
        \newline
        \includegraphics[height=1.7cm]{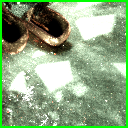}
        \newline
        \includegraphics[height=1.7cm]{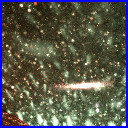}
    \end{minipage}
   \\
   MSE $\times 10^{2}$:&&\textbf{0.26}&0.95&1.11&1.06&0.29\\
   (1-SSIM)$\times 10^{1}$:&&\textbf{0.57}&1.38&2.22&1.95&0.82\\
   Time (s):&&31&39&26&27&33\\
\end{tabular}
\caption{Comparative rendering results of the \textsc{Ring} scene, featuring intricate and aesthetically appealing caustics cast by metallic rings and diamonds, with various photon guiding techniques under a fixed 256 iteration. The insets are isolated caustics for comparison purpose. (a) Our method leverages 3D Gaussian mixture to guide photon emission to improve the photon density in visible area, achieving accurate photon density estimation, with minimal overhead. (b) A na\"ive bound-based guiding (BG) cannot be optimized to fit the complex geometry. (c) 2D histogram (H2D) approach fails to fit accurate distribution for area lights due to parallax issue, yielding noisy result. (d) Similar to the 2D histogram (H2D) approach, the 3D von Mises-Fisher distribution (vMF) is also able to guide photon emissions directionally but encounters the same parallax issue, leading to suboptimal guiding efficiency. (e) State-of-the-art Markov chain Monte Carlo (MCMC) approach suffers from non-uniform convergence, leading to a higher occurrence of outliers and a slightly elevated overall error in comparison to our method. The reference is produced with the same integrated rendering system using uniform sampling for sufficiently many iterations.}
\label{fig:teaser}
\end{teaserfigure}

\maketitle

\section{Introduction}
In the realm of computer graphics, the rendering of caustics stands out for its visual appeal. In production rendering systems, caustics are usually separately sampled with photon mapping / photon tracing, while other components are path-traced. However, it requires significant photon density to render caustics accurately and effectively with photon mapping. This becomes a challenge when only a small portion of scene is rendered, or when only a selection of objects cast caustics. Many rendering systems thus necessitate photon guiding techniques to address this issue to achieve robust photon mapping.

There has been significant research focus on developing sufficient photon guiding techniques. \textit{Lightweight Photon Mapping}~\shortcite{LWPM} utilizes 2D histograms to learn distributions and guide emission for each light source. Hachisuka et al.~\shortcite{Hachisuka:MLT} leverage Markov chain Monte Carlo (MCMC) for more efficient photon emission sampling. However, limitations exist in these approaches. \textit{Lightweight Photon Mapping}'s 2D histogram approach is inherently limited by the parallax issue with non-delta light sources (e.g., rectangular light with nonnegligible area, or environment map with multiple peaks). MCMC occasionally suffers from non-uniform convergence, which is more visible when the rendering task consists of complex geometry and lighting setup. The limitations of existing methods drive us to seek a different approach to tackle the photon guiding problem.


In this paper, we propose a novel method to learn 3D \textit{spatial} distributions of photons emitted towards from each light source to maximize the photon density in the rendered image, which can greatly enhance the rendering performance of scenes with caustics.  
To achieve this objective, we first represent the 3D spatial distribution of the photons emitted to the environment from each light source by 3D Gaussians; we then derive a directional probability density function and method to sample directions from 3D Gaussian from any observation point, directly and unbiasedly, to guide photon emission.
This approach addresses and eliminates the common parallax issues found in directional distributions.
Our representation and derivation allow estimating the region to concentrate the emission of photons for increasing the density of photons at regions where visible caustics 
are produced.
To enhance the efficiency of online learning, we propose a scene-geometry-based initialization method that provides a precise starting point for the distribution. Along with 3D Gaussian emission guiding, we introduce a practical adaptive light source sampler. Together, these components form a comprehensive photon guiding framework. Our method is compact, executes rapidly on standard GPUs, and has been integrated with a unidirectional GPU path tracer to facilitate efficient rendering of caustics.

In conclusion, the contribution of this paper includes:

\begin{itemize}
    \item The development of an algorithm for sampling directions from a 3D Gaussian at any observation point, completed with a closed-form probability density function (PDF).
    \item A workflow to learn points of interest distribution for photon emission, represented with 3D Gaussian mixture, utilizing online photon samples as training data.
    \item A novel, robust photon guiding framework that progressively learns global distributions to guide photon emission, initializes the distributions with the scene geometry, and selects light sources with an adaptive sampler.
\end{itemize}

\section{Related work}
\subsection{Photon mapping and photon guiding}
Photon mapping is an efficient approach proposed by Jensen~\shortcite{Jensen:PM} to calculate global illumination. Later research proposed several modification on the scheme for more accurate density estimation and adaptation to progressive rendering scheme \cite{Hachisuka:SPPM, PPM}. Recently, Zhu et al.~\shortcite{DeepDensityEstimator} proposed to improve the quality of density estimation with machine learning. To address the biasedness of photon mapping, Qin \etal\shortcite{Qin:UnbiasedGathering} propose to generate another full light path at the photon position during gathering for unbiased result, and Misso et al.~\shortcite{Debias} showed that it is possible to obtain unbiased result from photon mapping with their debiasing framework. Photon mapping is considered as an effective approach to render caustics from specular surfaces such as glass and mirror. Even in hybrid methods which utilize multiple sample algorithms for robust rendering (e.g., \cite{VCM}), photon mapping is the common choice for caustics paths. Traditionally in photon mapping, photon emission is sampled uniformly; however, in practical scenarios, it is a common challenge that when only a small portion of the large scene can benefit from rendering caustics with photon mapping (the "caustic in a stadium" problem), the photon density with uniform emission sampling is not sufficient for accurate estimation. Hachisuka \etal\shortcite{Hachisuka:MLT} adapt Markov chain Monte Carlo (MCMC) for photon emission sampling, utilizing the primary sample space as introduced by Kelemen et al.~\shortcite{PSSMLT}. Grittmann et al.~\shortcite{LWPM} adapts 2D histogramming, to guide photon emission for higher photon density. These ideas are implemented in production renderers (e.g., \cite{Corona:Caustics, Hyperion, PathOfWater, Sony:PhotonGuiding}). However, we found these approaches to exhibit certain limitations: MCMC suffers from non-uniform convergence when the sample space is complex, while 2D histogram suffers from inherent parallax issue since it projects high dimensional distributions to 2D. Our method fits global 3D distributions, using a compact parametric model, to guide photon emission for light sources. The fitting process is efficient, and the model only uses several tens of scalars, providing a robust solution for the photon guiding problem.

\subsection{Distributions and sampling methods}
Our research seeks to importance sample photon emission with an explicit global distribution. In Monte Carlo rendering, it has become a common practise to importance sample different distributions for variance reduction. As a typical instance, almost every rendering system importance sample the local bidirectional scattering distribution function (BSDF) of surfaces (e.g., \cite{TrowbridgeReitz, OrenNayar}). However, sampling local BSDF distribution may not be optimal when indirect lighting is a major contribution to the result. Path guiding research leverage more general distributions to model indirect lighting. Jensen~\shortcite{JENSEN95} utilizes 2D histograms to model and importance sample the spherical incident radiance distribution. Many distribution models are explored subsequently. Vorba et al.~\shortcite{VORBA14} utilize 2D Gaussian mixtures for a similar purpose, with training data from a photon-emitting pass, to achieve a more practical method. Mueller et al.~\shortcite{PPG} propose an SD-tree to achieve more compact histogram-based representation. Rupper et al.~\shortcite{PAPG} propose to use von-Mises Fisher (vMF) distribution along with a parallax-aware fitting algorithm to remove the parallax issue caused by spatial partition. vMF is also utilized by Li et al.~\shortcite{SGCaustics} specifically for caustics rendering with unidirectional path tracing. Dodik et al.~\shortcite{SDMM} propose to use 5D Gaussians to model incident radiance distribution over the space. Huang et al.~\shortcite{NASG} propose an anisotropic spherical distribution that can be directly evaluated and sampled for more accurate and compact representation. 
Actually, despite different algorithms, these methods share a common insight, i.e. to build and sample local directional distributions that are represented by a suitable distribution model.

Previous photon guiding methods adapt this idea to photon emission straightforwardly. For example, several studies and rendering systems~\shortcite{LWPM,Hyperion,PathOfWater,Sony:PhotonGuiding} choose to build a local distribution for each of the light source using 2D histograms, however, as mentioned above, it suffers from the parallax issue due to the existence of area in actual light sources. Our research learns a per-light global distribution of the surfaces that photons should be emitted towards, and we find 3D Gaussian serves this purpose well. However, guiding photon emission from a light source with 3D Gaussians requires to sample directions from the distribution. To the best of our knowledge, there is no existing algorithm related to this. In this paper we derive a novel directional transform for 3D Gaussians, enabling sampling directions from it, to guide photon emission at any location. Reibold et al. \shortcite{Reibold:SelectiveGuiding} build unique guiding distributions using truncated Gaussians on-the-fly from previously collected challenging light paths for each sampling event. This method shares a similar adaptive philosophy with our approach.

In addition to various distributions used in importance sampling, Markov chain Monte Carlo (MCMC) methods, originating from the pioneering work by Metropolis et al.~\shortcite{Metropolis:MCMC}, have been extensively studied in Monte Carlo rendering. Veach and Guibas~\shortcite{Veach:MLT} introduced Metropolis Light Transport (MLT), a seminal technique that utilizes MCMC to efficiently sample complex light paths, significantly improving the rendering of global illumination. Building on this foundation, Kelemen et al.~\shortcite{PSSMLT} developed primary sample space MLT, which enhanced the robustness and efficiency of the sampling process. Kitaoka et al.~\shortcite{Kitaoka:ReplicaExchange} further extended these concepts with the replica exchange light transport method, combining MCMC with replica exchange to tackle challenging light paths more effectively. These foundational works underpin state-of-the-art MCMC photon emission techniques.

\subsection{Adaptive light source sampling}
In addition to emission guiding, it is also important to importance sample light sources. Traditionally light sources are sampled based on their flux, however adaptive sampling strategy is even preferred as reported in \cite{Corona:Caustics}, since the actual contribution of each light source does not (fully) depend on their flux when only a portion of the full scene is rendered. Walter et al.~\shortcite{LightCut} propose Lightcuts, which utilizes a binary tree, to cluster light sources for efficient rendering. Yuksel~\shortcite{StochasticLightcuts} sample Lightcuts stochastically in Monte Carlo rendering to remove bias. Wang et al.~\shortcite{ProgressiveLightSampler} propose to adaptively refine the importance of each node of the binary tree. Our adaptive light sampler can be seen as an adaptation of the idea of Wang et al. to photon guiding, with the difference that we progressively build the binary tree instead of starting from complete Lightcuts, and we approximate the importance of each node based on the sum of radiance from gathered photons.
\section{Overview}

\begin{figure*}[h]
    \centering
    \includegraphics[width=0.7\linewidth]{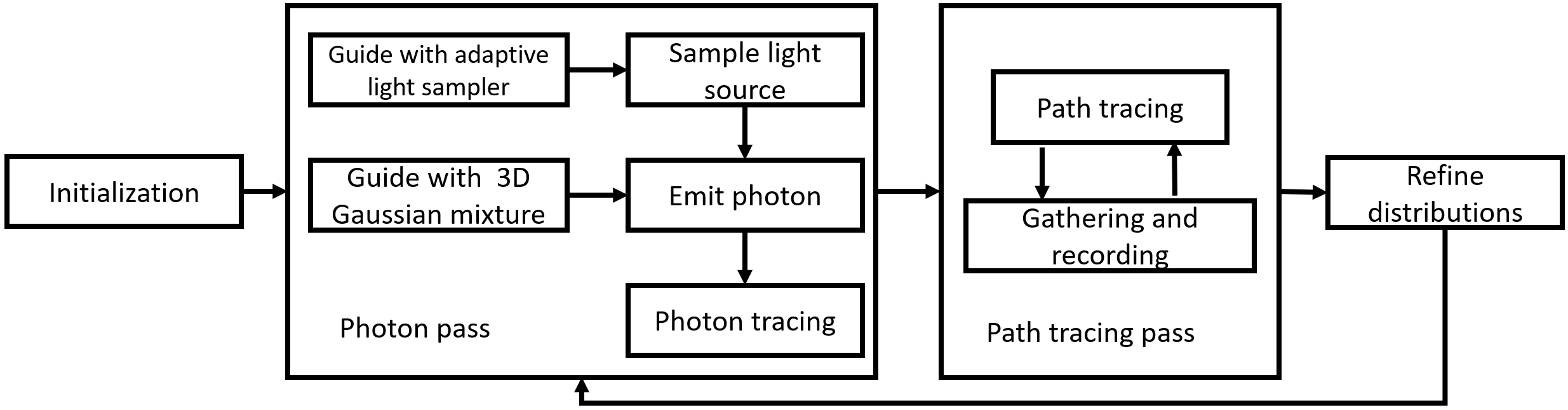}
    \caption{Our method is integrated in a two-pass rendering process. After the photon pass and the path tracing pass, the distributions are refined using the collected photon data recorded in path tracing pass. With guiding of the learned distributions, the photon density gradually increases over iterations.}
    \label{fig:workflow}
\end{figure*}

\paragraph{Background} Our technique improves photon density for visible region to achieve higher quality photon mapping results. To better explain our technique with an actual implementation, we first briefly describe the approach we integrate photon mapping into a production renderer for caustics rendering. Since the target production renderer is a GPU unidirectional path tracer, we employ photon mapping to render caustics in stochastic approach (i.e., \cite{Hachisuka:SPPM}), allowing the rest to be handled by path tracing. In the target path tracer, the specular surfaces that casts caustics are explicitly marked as ``casters'' while surfaces that receives caustics are marked as ``receivers''. Only photons that travels through casters and reach receivers will be recorded to construct the photon map. This is a typical approach in production rendering systems, however, under this setup, uniform emission cannot achieve significant photon density for proper density estimation, leading to poor results. Therefore, we propose to guide photon emission with a novel online learned distribution.
\paragraph{Integration} \figref{workflow} outlines the components and workflow of our framework. The main rendering pipeline is a two-pass process: in the first pass (the photon pass), we randomly trace a batch of photons, guided by our progressively refined distributions, and build a photon map with a KD-tree. The second pass is path tracing; in addition to regular path tracing, we gather nearby photons and estimate indirect lighting at diffuse surfaces. During gathering, we also record the data required for distribution learning. After each rendering iteration, the distribution is refined based on the recorded data. Through the rendering iterations, our distribution converges to one that maximizes the number of gathered photons. The distribution consists of two parts: a 3D Gaussian mixture that represents the spatial distribution of points of interest, which is used for guiding later photon emission from light sources (see \secref{G3D}); and a tree-based 1D distribution that is used for light source sampling (see \secref{ALS}).

\section{Learning to guide photon emission with 3D Gaussians}\label{sec:G3D}
In this section, we briefly review the existing photon guiding approaches, as well as other potential methods that could be applied yet not explored. Afterwards, we introduce our approach to guide photon emission with 3D Gaussians.

\subsection{Existing techniques}\label{sec:Existing}
\paragraph{2D Histograms} For most of the light sources, the photon emission can be seen as a 4D distribution $D(x,y,\theta,\phi)$ where $x,y$ defines the point at the light surface and $\theta, \phi$ represents the direction. However, since most of the light sources that cause caustics only have a small area, previous photon guiding methods choose to only guide the direction components. Grittmann \etal \shortcite{LWPM} leverage 2D maps to histogram the 2D directional distribution for guiding. It projects $\theta$ (usually it's $\cos\theta$ for better sampling efficiency) and $\phi$ to a 2D square with a fixed resolution. During emission sampling, the coordinate is sampled from the 2D map based on the pixel weights and converted to directions. Droske \etal\shortcite{PathOfWater} further improve this idea by introducing a quadtree representation as a replacement of fixed-resolution 2D histograms; the quadtree is progressively subdivided into smaller squares at areas with high weight, so that the resolution is not limited.

The directional distribution often encounters a parallax issue when modeling area lights: it assumes emissions originate from a single point. This simplification means the fitted distribution may not accurately reflect the true conditions when applied at specific points other than the assumed position, leading to inaccuracies in the rendered scene.

\paragraph{von Mises-Fisher mixture}
In computer graphics and statistics, 3D von Mises-Fisher (vMF) distribution has many applications:
\begin{equation}
    F_v(\omega~|~\kappa, \nu) = \frac{\kappa}{4\pi\sinh(\kappa)} e^{\kappa (\cos\theta-1)},
\end{equation}
where $\theta$ is the angle between $\omega$ and vMF's mean vector $\nu$: $\cos\theta = \omega \cdot \nu$. The application of vMF in photon guiding is not explored yet, however, it can easily work as a replacement for 2D histograms for direction guiding. There is also normalized anistropic spherical Gaussian (NASG \cite{NASG}) that serves a similar purpose with anisotropic support. Such spherical parametric distributions share the same limitation with 2D maps: it cannot guide infinite light sources, and it suffers from the parallax issue.

\paragraph{Bounding box}
Since our primary interest lies in guiding photon emissions toward the casters, an alternative approach is to leverage the bounding boxes of the casters. These bounding boxes can be interpreted as a spatial distribution, transformable into a directional distribution from any point, thereby avoiding the parallax issue. In rendering systems, axis-aligned bounding boxes (AABB) are typically available with no additional computational cost since it's also widely used in other aspects. There are also alternatives to AABB, such as oriented bounding box (OBB) and bounding sphere.

While the selection probability for each bounding box can be optimized during the rendering process, the distribution inside a bounding box cannot be further refined. Contrary to the assumption of an even distribution within bounding boxes, the actual distribution is typically non-uniform. This discrepancy reduces the effectiveness of bound-based guiding, particularly when only a small part of the bounding area contributes to visible caustics—common in situations like a close-up view of underwater caustics or when surfaces are unevenly distributed within bounds (e.g., within a ring). Furthermore, in scenes with numerous casters, the process of multiple importance sampling across many bounds can become excessively costly.
\subsection{3D Gaussians}
We propose to use 3D Gaussians for photon guiding to overcome the major limitations discussed in \secref{Existing}. 3D Gaussians can be optimized easily, and can be applied to model spatial distributions that does not suffer from parallax issues. We will also show that with a trivial transformation, we can sample directions directly from 3D Gaussians.

\begin{figure}[h]
\setlength{\abovecaptionskip}{2pt}
    \centering
    \includegraphics[width=0.9\linewidth]{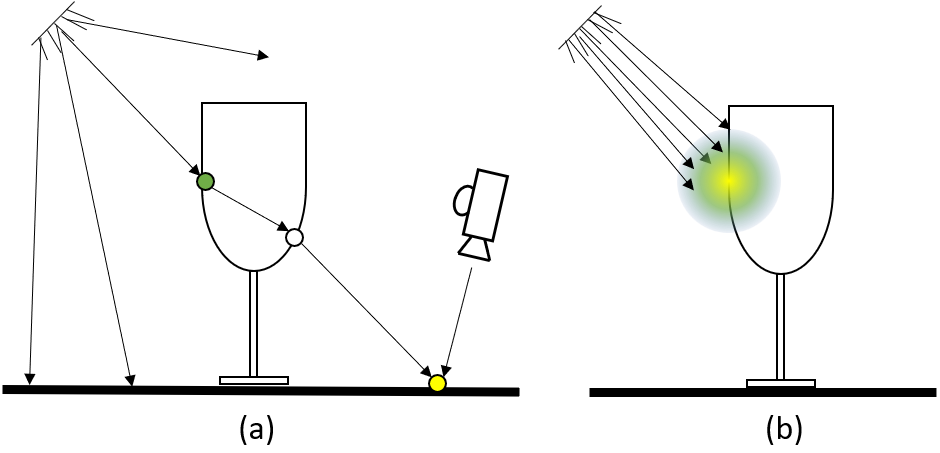}
    \caption{The concept of photon guiding with 3D Gaussians: we fit a 3D Gaussian mixture with gathered photons and use it to guide photon emission. In (a), the light source uniformly emits multiple photons, however, only the yellow photon is actually gathered from the camera view. We use the first bounce location of yellow photon to fit a 3D Gaussian as shown in (b). Later when emitting photons, we use the 3D Gaussian to sample emission directions, so that more photons can be gathered, achieving higher density.}
    \label{fig:g3d_fit_diagram}
\end{figure}

An (unnormalized) isotropic 3D Gaussian is given by the equation:
\begin{equation}
    G(\mathbf{x}) = \exp\left(-\frac{\|\mathbf{x} - \mathbf{\mu}\|^2}{2\sigma^2}\right),
    \label{eq:g3d}
\end{equation}
where \( \mathbf{x} \) is the position vector in 3D space, \( \mathbf{\mu} \) is the mean vector representing the center of the Gaussian distribution and \( \sigma \) is the standard deviation.
A 3D Gaussian mixture with $N$ isotropic 3D Gaussians can effectively model distributions over the 3D space:
\begin{equation}
M(\mbx) = \sum_{i=1}^{N} \frac{w_i}{\sqrt{(2\pi\sigma^2)^3}}  G_i(\mbx),    
\end{equation}
where $G_i$ are the 3D Gaussian components, and $w_i$ are the corresponding weights which sum to 1. The normalizing constant is the integral of each $G_i$.
As shown in Fig. \ref{fig:g3d_fit_diagram}, for each light source, our method fits a global 3D Gaussian mixture, which is used to guide the later photon emission. 

In the rest of this section we describe the workflow to fit the 3D Gaussian mixture using gathered photons, and the method we derived to sample emission direction from a 3D Gaussian.

\subsection{Learning with gathered photons}

Our goal is to model a distribution to achieve maximum photon density in the visible region, and eventually improve the accuracy of density estimation. Based on the fact that caustics photons are usually gathered at a diffuse surface after a sequence of specular bounce, we only need to fit a distribution at the location of the first bounce, which we define as the point of interest. To learn the distribution of points of interest, when emitting photons, we record the location of the first bounce and the emission pdf (directional pdf $\cdot$ positional pdf), along with the original photon data. In addition, every photon has an indicator of how many times it is gathered.

We learn the distribution with gradient descent using Kullback-Leibler divergence (KL divergence), adapting the approach proposed in \cite{NIS}. First, we define the learned 3D Gaussian mixture as $q(\mbx;\gamma)$, where $\gamma$ is the vector of parameters. The target distribution $p(\mbx)$, which we are trying to fit, is unknown. However, for each point of interest, the basic rule is that the more times the photons are gathered the higher the density should be, thus we define:
\begin{equation}
    p(\mbx) = g\cdot T(\mbx),
\end{equation}
where $T(\mbx)$ denotes how many times a photon bounced from $\mbx$ is gathered in the current pass, and $g$ is an unknown global scaling factor. The fitting can then be achieved by minimizing a KL divergence: 
\begin{equation}
D_{KL}(p(\mbx)\|q(\mbx;\gamma)) = \int_{\mathbb{R}^3}p(\mbx)(\log[p(\mbx)] - \log[q(\mbx;\gamma)])d\mbx.
\end{equation}
We can minimize it by using gradient descent by optimizing \(\gamma\), where \(\gamma = (\mu, \sigma)\) in our context. Notice that $\log[p(\mbx)]$ is irrelevant to $\gamma$'s gradient and thus we have
\begin{equation}
	\nabla_{\gamma} D_{KL}(p(\mbx)\|q(\mbx;\gamma)) = -\nabla_{\gamma}\int_{\mathbb{R}^3}p(\mbx) \log[q(\mbx;\gamma)])d\mbx.
\end{equation}
Although it requires to calculate the integral over 3D space, we are able to attempt a one-sample estimation:
\begin{equation}
\nabla_{\gamma} D_{KL}(p(\mbx)\|q(\mbx;\gamma)) = -\, \mathbb{E} \left[ \frac{p(\mbx)}{\hat{q}(\mbx)}\nabla_{\gamma}\log[q(\mbx;\gamma)]\right].
\end{equation}
In our rendering process, each time we sample an emitted photon, the emission pdf is $\hat{q}(\mbx)$. We then replace $p(x)$ with $g\cdot T(\mbx)$, and by using a moment-based optimizer, the global scaling factor $g$ can be effectively cancelled.
\subsection{Sampling directions from 3D Gaussian} \label{sec:SamplingG3D}
In this section, we describe a novel approach to transform the a global 3D Gaussian mixture that represents the distribution of photons, into a local directional distribution representing the photon emission from a given light source position.      

Given an observation point $\mbx_0$, we can define the north pole $\mbz$ pointing from $\mbx_0$ to $\mu$. Then, we are able to represent any 3D point $\mbx$ using distance $r = \|\mbx_0-\mbx\|$ and the angle $\theta$ between $(\mbx - \mbx_0)$ and $\mbz$:
\begin{gather}
    \mbz = \frac{\mu-\mbx_0}{d} \text{ where } 
    d = \|\mathbf{x_0} - \mathbf{\mu}\|.
\end{gather}
Then, $\|\mbx-\mu\|^2$ in \eqref{g3d} can be represented with distance $r = \|\mbx_0-\mbx\|$ and $\theta$:
\begin{equation}
    \lVert \mathbf{x} - \mu \rVert^2 = \lVert (\mathbf{x} - \mathbf{x}_0) - (\mu - \mathbf{x}_0) \rVert^2 = (r - d\cos\theta)^2 + d^2\sin^2\theta.
\end{equation}
With these parameters, we transform a 3D Gaussian to polar coordinates:
\begin{equation}
    G(\omega, r) = \exp\left(-\frac{(r - d\cos\theta)^2 + d^2\sin^2\theta}{2\sigma^2}\right).
\end{equation}
The integral of 3D Gaussian can be then transformed to the spherical form:
\begin{equation}
\label{eq:directional_integral}
    \int_{\mathbb{R}^3} G(\mathbf{x}) \, d\mathbf{x} =\int_\Omega \int_0^{\infty} \exp\left(-\frac{(r - d\cos\theta)^2 + d^2\sin^2\theta}{2\sigma^2}\right) r^2 dr d\omega.
\end{equation}
By matching \eqref{directional_integral} with the definition of a directional distribution, we derive the directional distribution of a 3D Gaussian at observation point $\mbx_0$ as:
\begin{equation}
f_o(\omega) = \int_0^{\infty} \exp\left(-\frac{(r - d\cos\theta)^2 + d^2\sin^2\theta}{2\sigma^2}\right) r^2 dr.
\end{equation}
Actually, this integral has a closed-form solution:
\begin{equation}
\begin{split}
f_o(\omega) = & \, \sigma^2 e^{-\frac{d^2}{2\sigma^2}} d\cos\theta \\
& + \sqrt{\frac{\pi}{2}} \sigma e^{-\frac{d^2 \sin^2\theta}{2\sigma^2}} \left( \sigma^2 + d^2 \cos^2\theta \right) \\
& \times \left( 1 + \text{erf}\left( \frac{d \cos\theta}{\sqrt{2}\sigma} \right) \right).
\end{split}
\end{equation}
This is our unnormalized directional PDF of the 3D Gaussian; readers are referred to the appendix for a complete derivation. Since the normalizing term is the integral of the original 3D Gaussian over real space, the normalized form is given as:
\begin{equation}
    F_o(\omega) = \frac{1}{\sqrt{(2\pi\sigma^2)^3}}f_o(\omega).
\label{eq:directional}
\end{equation}
The distribution can be effectively evaluated; it only involves several elementary transcendental functions and the error function, and the error function can be precisely approximated with \textit{Abramowitz and Stegun approximation}~\shortcite{abramowitz_stegun}. Despite the complex form, we find it rather simple to sample this distribution too: by sampling a 3D point from the original 3D Gaussian, the distribution of direction pointing to it from the observation point obeys \eqref{directional}. In the appendix we provide the complete proof of this sampling algorithm. With these conditions, we achieve efficient unbiased sampling of spherical transform of 3D Gaussian.

Our transformation of the 3D Gaussian essentially builds a connection between global spatial distribution and local directional distribution; this can potentially lead a shift from using multiple local directional distribution (e.g., the vMF mixture) to one global 3D Gaussian mixture, which could significantly reduce redundancy and improve learning efficiency.
Furthermore, it could eliminate the parallax issue introduced by local discrete distributions.

\begin{figure}
\setlength{\abovecaptionskip}{1.5pt}
\setlength{\belowcaptionskip}{1pt}
    \centering
    \begin{tabular}{@{\hspace{-5pt}}c@{\hspace{0pt}}c@{\hspace{-5pt}}}
        \includegraphics[width=0.5\linewidth]{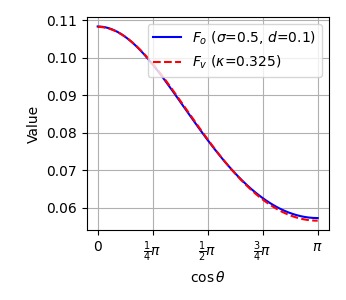} &\includegraphics[width=0.5\linewidth]{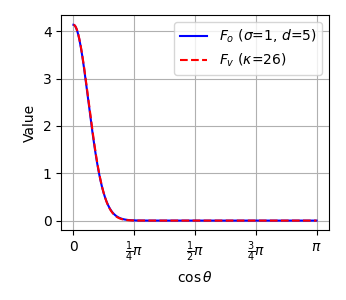} 
    \end{tabular}
    \caption{Visualization of distributions for our directional transform of 3D Gaussian ($F_o$) and vMF ($F_v$) with different parameters. By choosing parameters meticulously, the two distributions achieve very similar shapes.}
    \label{fig:distribution_vis}
\end{figure}

\paragraph{Similarity with von-Mises Fisher distribution}$d$ and $\sigma$ determine the shape of $F_o(\omega)$. Interestingly, we find the shape of this spherical distribution very similar to 3D von-Mises Fisher distribution (vMF).
Actually, the north pole $\mbz$ used in \eqref{directional} works the same way as the mean vector $\nu$ of vMF (thus, in our discussion, we always assume $\mbz$ and $\mathbf{\nu}$ are the same). 
\figref{distribution_vis} presents a comparative visualization: it displays \( F_o(\omega) \) under two distinct parameter sets alongside a vMF distribution. The parameters of the $F_v$ are meticulously chosen to achieve a close match with the shape of \( F_o(\omega) \). 

\subsection{Sampling Photon Emission with 3D Gaussian Mixture}
Photon emission sampling comprises two principal stages: determining the emission location $\mbx_0$, and selecting the emission direction $\omega_0$. Utilizing 3D Gaussians, we can precisely guide photon emissions from different types of light sources, each following a unique approach:
\begin{description}
\item[Area and Point Lights:] For both area and point lights, we first sample the emission location $\hat{\mbx}_0$. In the case of area lights, we perform this sampling uniformly across the surface. For point lights, the location is a fixed point. Subsequently, we determine the emission direction using the method we describe in \secref{SamplingG3D}.
\item[Infinite Lights:] We begin by sampling the direction $\hat{\omega}_0$ from the infinite light's distribution. A plane is then defined that passes through the center of the scene and is perpendicular to $\hat{\omega}_0$. Projecting our 3D Gaussian mixture onto this plane yields a corresponding 2D Gaussian mixture. We sample a point $\hat{\mbx}_0'$ from this 2D distribution on the plane. The actual emission location $\hat{\mbx}_0$ is calculated as $\hat{\mbx}_0 = \hat{\mbx}_0' - b\hat{\omega}_0$, where $b$ is the radius of the scene's bounding sphere. Importantly, the 2D Gaussian has infinite support, and truncation to the bounds of the scene's bounding sphere is not necessary.
\end{description}
To sample from a 3D Gaussian mixture, we first select a component based on its weight and sample the projected direction or 2D position from it. The probability density function (PDF) for each sample is then calculated as the weighted sum of the PDFs from all components at that sample point.
\section{Photon Guiding Framework}
To enhance the efficiency and accuracy of photon guiding, our method incorporates two pivotal components alongside the 3D Gaussians optimization and sampling. These components, a robust initializer and an adaptive light sampler, are designed to significantly improve the learning process and photon distribution. In this section, we elaborate on each of these components, detailing their implementation and contributions to the overall framework.
\subsection{Robust Initializer}
In line with standard practices in distribution-fitting applications, our 3D Gaussian mixture model requires an accurate initial guess to commence the fitting process effectively. 

A na\"ive approach can utilize the first batch of photon emission: for a 3D Gaussian mixture comprising \( N \) components, we employ the k-means method to ascertain \( N \) mean positions from the \textit{initial position} data in the first photon pass. Subsequently, these mean positions serve as the preliminary guesses for the Gaussian components. Regarding the standard deviation parameter (\( \sigma \)), we opt for the median value of our encoding.

This can be further extended to a more robust initialization that leverages the prior knowledge of scene geometry. Before the rendering, we fit a global 3D Gaussian set from the caustics casters' geometry data using the k-means method. After the first emission batch, for each light source, we select the 3D Gaussians that the first hit of the photons are closest to, and pick the Top N selected Gaussians as the initial set for that light source. The detailed algorithm is described as Algo.~(\ref{alg:initializer}). Compared with the na\"ive approach, our initialization ensures fast convergence. Note that such initialization is not trivial with directional mixture models such as vMF.

\begin{algorithm}
\caption{Scene-geometry-based 3D Gaussian Mixture initializer}
\label{alg:initializer}
\begin{algorithmic}[1]

\Require Scene geometry list $S$, number of Gaussians per geometry $K$, number of photons $P$, number of components $G$, light source list $L$

\Function{initializeLightSourceGaussians}{$S$, $K$, $L$, $G$}
    \Statex \textit{\textcolor{blue}{// Create and initialize Gaussians from scene geometry}}
    \State $\textit{gaussians} \gets \text{initializeGaussians}(S, K)$
    
    \Statex \textit{\textcolor{blue}{// Emit photons and gather them}}
    \State $\textit{photons} \gets \text{emitPhotons}(P)$
    \State $\textit{gatheredPhotons} \gets \text{gatherPhotons}(\text{camera}, \textit{photons})$
    
    \Statex \textit{\textcolor{blue}{// Compute initial Gaussians for each light source}}
    \ForAll{$l$ in $L$}
        \State $\textit{Gs} \gets \textit{gaussians}, \textit{GPs} \gets \textit{gatheredPhotons}$
        \State $\textit{topG} \gets \text{computeInitialGaussians}(\textit{Gs}, \textit{GPs}, G)$
        \State $\text{setLightSourceGaussians}(l, \textit{topG})$
    \EndFor
\EndFunction

\Function{computeInitialGaussians}{$gaussians$, $photons$, $G$}
    \Statex \textit{\textcolor{blue}{// Increment counter for closest Gaussian to each photon}}
    \ForAll{$g$ in $gaussians$}
    \State $\text{clearCounter}(g)$
    \EndFor
    \ForAll{$p$ in $photons$}
        \State $g_{\text{closest}} \gets \text{findClosestGaussian}(\textit{gaussians}, p)$
        \State $\text{incrementCounter}(g_{\text{closest}})$
    \EndFor
    \Statex \textit{\textcolor{blue}{// Sort Gaussians by counters and select top $G$}}
    \State $\text{sortGaussiansByCounter}(\textit{gaussians})$
    \State $\textit{topGaussians} \gets \text{selectTopG}(\textit{gaussians}, G)$
    \State \Return $\textit{topGaussians}$
\EndFunction

\end{algorithmic}
\end{algorithm}


\subsection{Adaptive Light Sampler}\label{sec:ALS}
Under the circumstance that only a portion of objects cast caustics, an independent light source sampler helps to reduce the variance and improve photon density. \cite{Corona:Caustics} mentioned a light source sampler for photon mapping is achieved, however the authors didn't provide details of the method. In this section we introduce our adaptive light sampler, which progressively learns an appropriate distribution of light sources for photon emission.

Our light sampler is inspired by \cite{ProgressiveLightSampler} and \cite{PPG}: we store a binary tree whose nodes are used to store the photon count gathered in the last iteration. Each node represents a range of light sources according to the node's position and depth. For example, the root node represents the whole light sources, while the left child of it represents the first half. Starting from a single root node, in each sample iteration, we record how many photons are gathered in the region a node represents. When the photon number of a leaf node passes the threshold, we branch the leaf node into two children. To sample this light tree we just need to run binary sample down along the tree, using each node's recorded count as importance. After several iterations we are able to obtain a close approximation of importance distribution of light sources. Compared to a 1D piecewise constant table approach, our adaptive sampler is efficient to be built in a streamed GPU renderer, and can quickly prune to important light sources when there are many light sources in the scene.

\section{Implementation Details}
We implement our method in a GPU production renderer. In this section we provide implementation details that help to build a complete learning-based photon guiding framework.
\paragraph{Encoding Scheme}
Optimizing (sparse) 3D Gaussians in large scenes presents significant challenges. In our implementation, we consistently apply a scaling factor \( B \) to the scene positions for optimization or guiding purposes. The scaling factor \( B \) is defined as:
\begin{equation}
    B = \frac{c_b}{R_b},
\end{equation}
where \( R_b \) is the diameter of the scene's bounding sphere, and \( c_b \) is a constant meta-parameter. Furthermore, we encode the \(\sigma\) of the 3D Gaussian using the sigmoid function to ensure robust and efficient optimization:
\begin{equation}
    \sigma = \frac{c_s}{1 + e^{-p_{\sigma}}},
\end{equation}
where \( p_{\sigma} \) is the learned encoded parameter, and \( c_s \) is a scaling meta-parameter. It is crucial to set \( c_b \) and \( c_s \) in a correlated manner to achieve stable optimization. In our implementation, we use \( c_b = 20 \) and \( c_s = 0.65 \). The weights of components in a mixture are normalized using softmax:
\begin{equation}
     w_i = \frac{e^{w'_i}}{\sum_{j=1}^N e^{w'_j}},
\end{equation}
where $w'_{i}$ is the learned unnormalized weight.
\paragraph{Progressive learning of 3D Gaussian mixture}
Although we use gradient-based learning approach to fit the global 3D Gaussian mixture, we do not use existing learning frameworks. Instead, we implement a high performance learning kernel utilizing auto-differentiation feature of \textsc{LuisaRender}~\shortcite{luisa}, which can be integrated with our rendering system's pipeline seamlessly. The learning process requires a moment-based optimizer, for which we utilize Adam \shortcite{Adam}. The learning rate is initially set to 0.1 and then progressively scaled down to 0.01.
\paragraph{Searching radius}
The searching radius is an essential parameter for gathering. We employ an adaptive searching radius approach, where a relatively large maximum radius is initially set. For each shading point, we gather the four nearest photons, and the distance to the farthest photon among these dictates the actual searching radius. This allows for dynamic adjustment of the radius in response to the increasing photon density achieved through photon guiding.
\paragraph{Multiple importance sampling with uniform emission}
Existing path guiding techniques usually blend guiding with uniform sampling: a selection probability $\beta$ is used to choose the distribution to sample from, and $\beta$ progressively increases from 0 to a fixed or learnable value. However, due to our robust scene-geometry-based initializer, we can directly use a high fraction from the beginning. In our implementation, our first rendering iteration uses uniform sampling and the result is only used for initialization. We discard this result, initialize the 3D Gaussians for each light source, and set $\beta$ to 0.8 in the rest of the iterations. 3D Gaussian has infinite support, which in theory ensures unbiased sampling.

\section{Experiments}
\begin{figure*}[h]
    \begin{tabular}{cccc}
    \includegraphics[height=2.0cm]{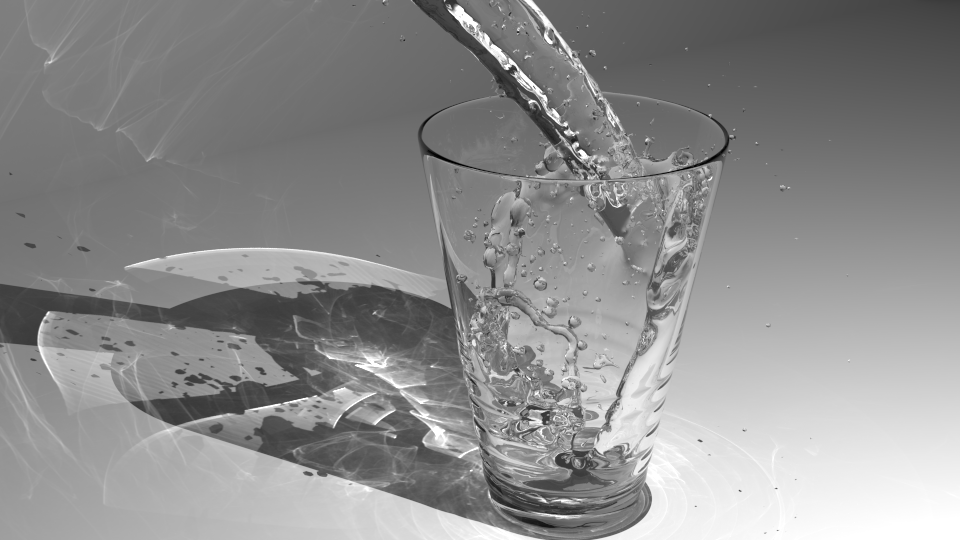}
    &
    \includegraphics[height=2.0cm]{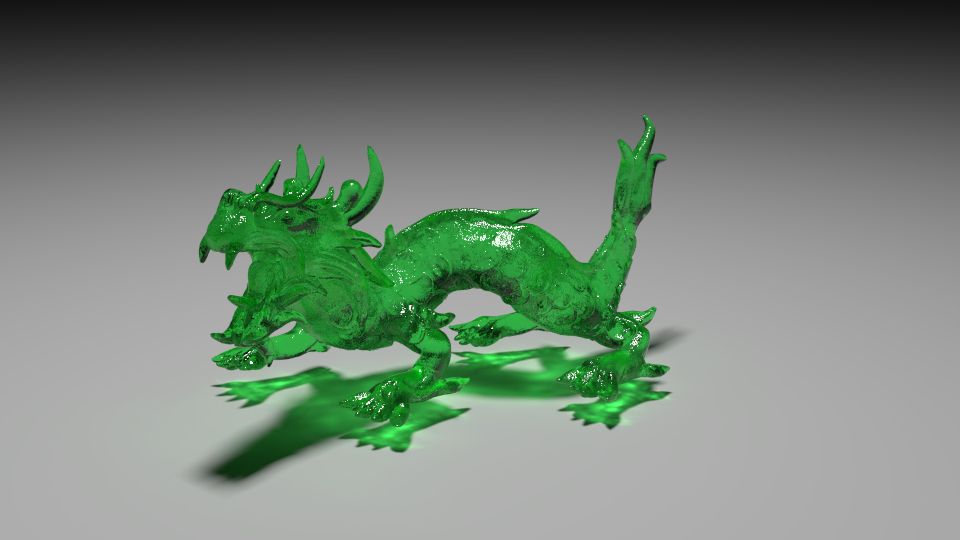}
    &
    \includegraphics[height=2.0cm]{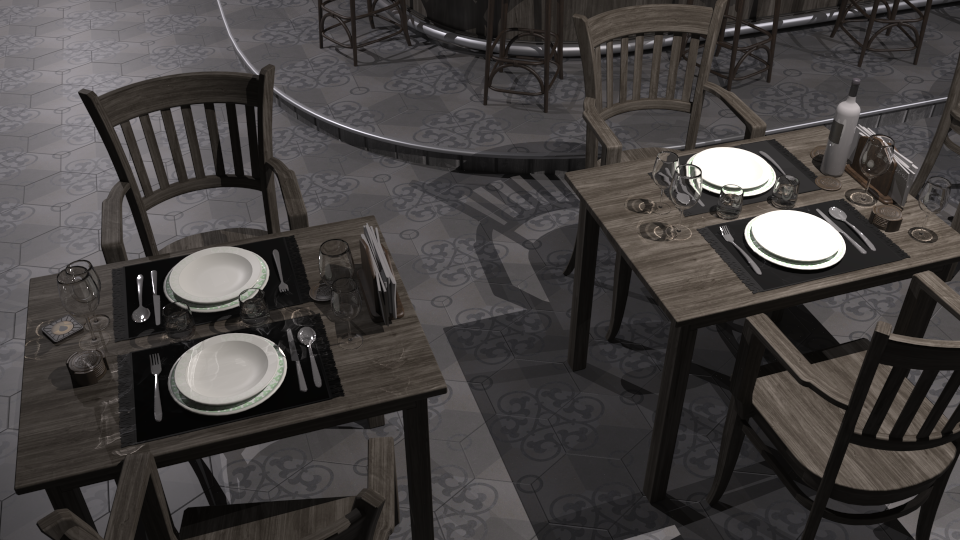}
    &
    \includegraphics[height=2.0cm]{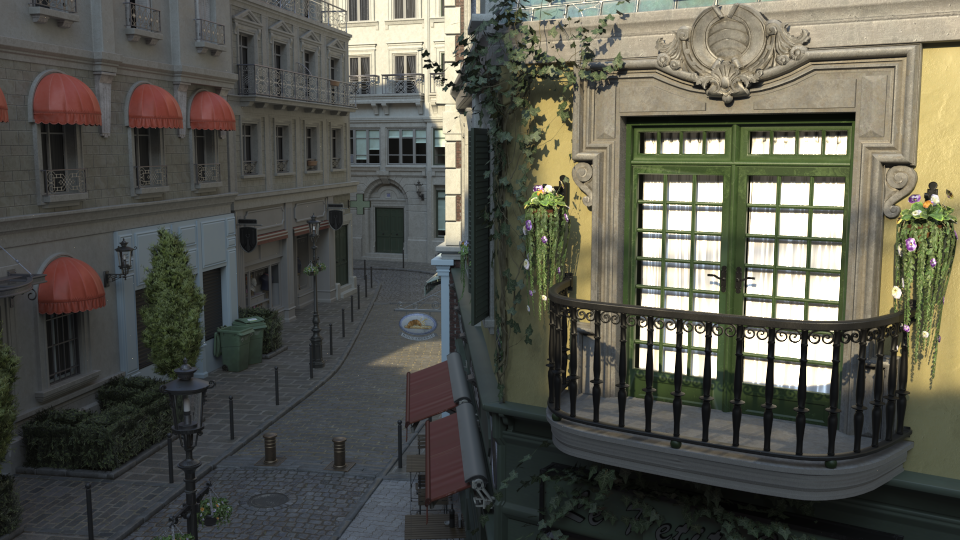}
    \\
    \includegraphics[height=2.0cm]{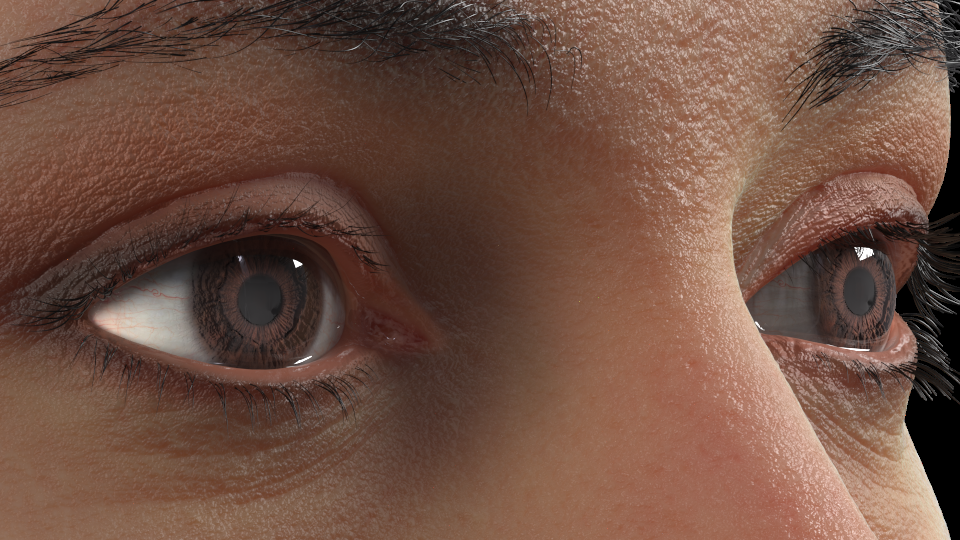}
    &
    \includegraphics[height=2.0cm]{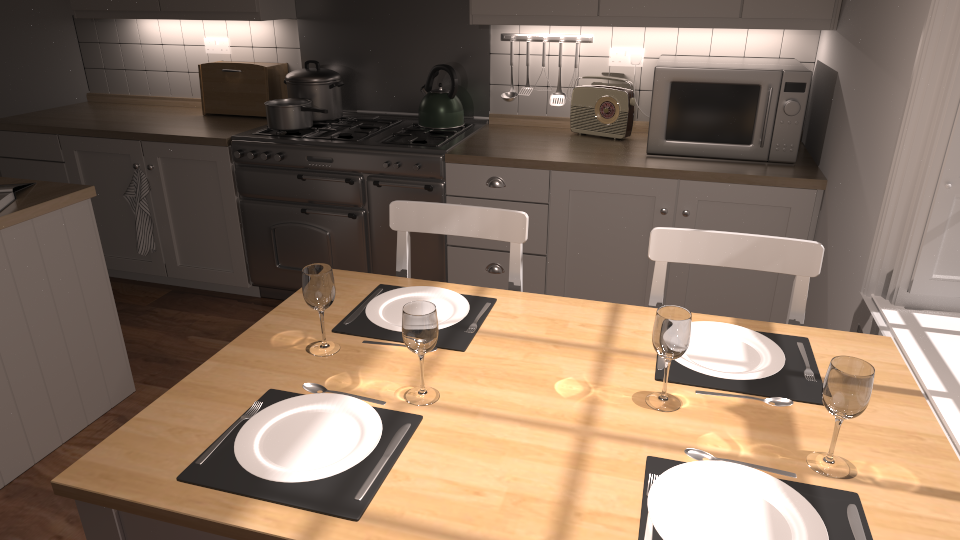}
    &
    \includegraphics[height=2.0cm]{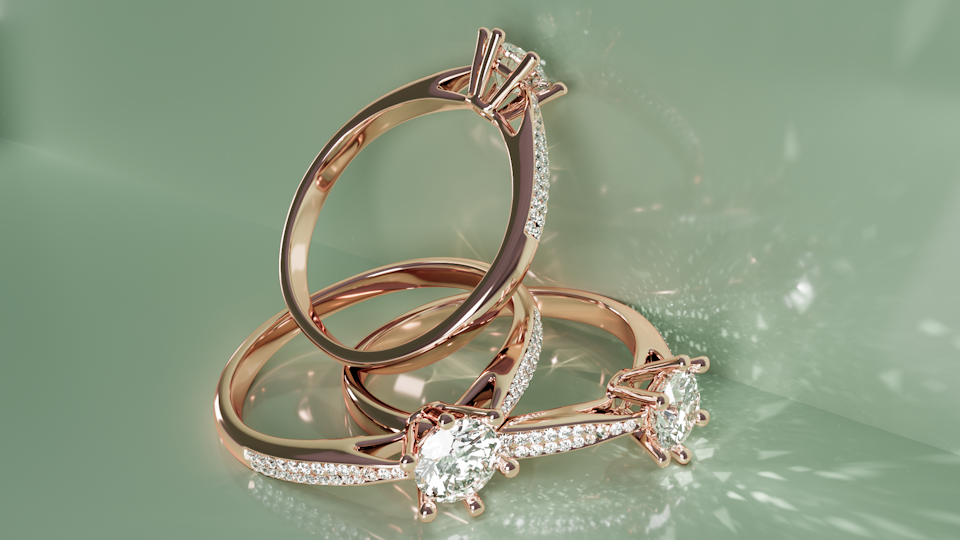}
    &
    \includegraphics[height=2.0cm]{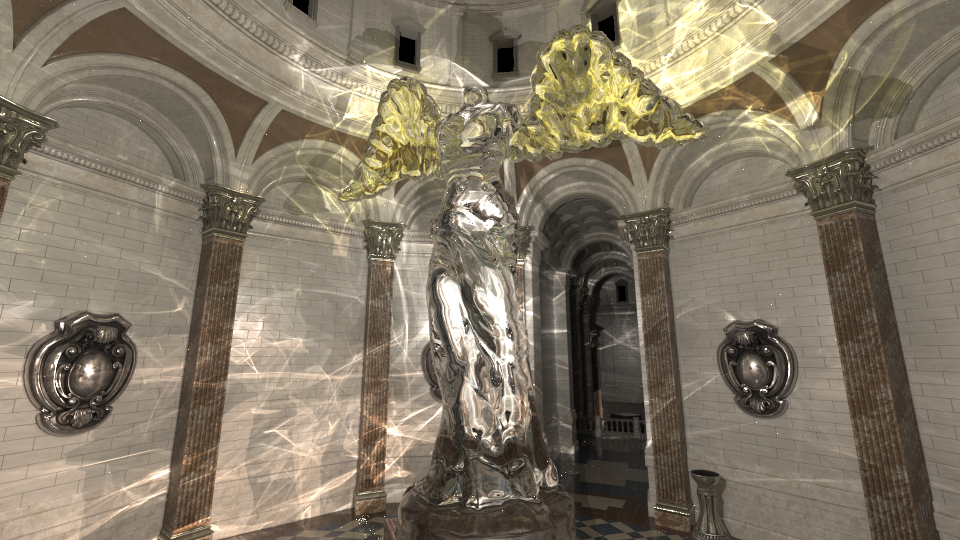}
    \end{tabular}
    \caption{The scene set we used for experiments: \textsc{GlassOfWater}, \textsc{Dragon}, \textsc{Bistro}, \textsc{BistroStreet}, \textsc{Eyes}, \textsc{Kitchen}, \textsc{Ring}, \textsc{SunTemple}.}
    \label{fig:scene_set}
\end{figure*}

We conduct evaluation experiments by rendering scenes using different configurations and techniques, and then comparing the quality of the outputs. All images in this section are rendered at a resolution of 1920$\times$1080, and full-sized images are included in the supplemental materials. The rendering is performed on the same conventional PC equipped with an Intel Core 9700K CPU and an Nvidia RTX 4070 GPU. In all experiments, our method utilizes 32-component 3D Gaussian mixtures. 
We use the same maximum searching radius for all methods (while keeping the adaptive strategy as described above). The output from the renderer includes all light paths; however, to evaluate the guiding quality, we specifically isolate the caustics results for comparison. The scenes we used are shown in \figref{scene_set}. These include simple caustic cases as well as complex, production-level assets and scenes that present challenges for photon mapping. We report the mean square error (MSE, where lower is better) and the complement of the structural similarity index measure (1-SSIM, where lower is better) for all comparisons to evaluate both image error and visual quality. These metrics are calculated using images rendered with uniform sampling through sufficiently many iterations as the reference. In our integrated renderer, photon gathering is the most computation-intensive stage. This means that methods emitting a higher number of visible photons are generally slower. Therefore, even though similar time costs are recorded across all methods, we do not compare them directly in our experiments due to the irrelevance of these differences in this context. Nonetheless, we include time cost data for the sake of completeness.

\subsection{Distributions Comparison} \label{sec:distributions}
In this experiment, we compare our 3D Gaussian (G3D) guiding with other existing distribution models, including 2D histograms (H2D), 3D von Mises-Fisher (vMF), AABB-based guiding (BG), and traditional uniform distribution (Uni). Although our method provides a full framework for photon guiding with 3D Gaussians, for fair comparison, we set the configurations as follows to achieve similar conditions for each distribution: 
\begin{itemize}
    \item To learn an accurate distribution, H2D uses a fixed resolution of $256\times256$ for all light sources, vMF and G3D use 32 components.
    \item Scene-geometry-based initializer is disabled and 3D Gaussians are initialized na\"ively.
    \item The selection probability $\beta$ starts from 0 and linearly increases to 0.75 in 128 iterations for G3D, H2D, and vMF. Since BG does not require any optimization, we do not blend BG with uniform sampling. However, we optimize the weights of each bound during rendering.
    \item Our adaptive light sampler is enabled for all the distributions. 
\end{itemize}

\begin{figure*}[h]
\setlength{\abovecaptionskip}{2pt}
    \begin{tabular}{crcccccc}
        &&Reference&G3D&BG&H2D&vMF&Uniform
        \\
        \rotatebox[origin=c]{90}{(a) \textsc{GlassOfWater}}&
        \begin{minipage}{0.3\linewidth}
            \centering
            \includegraphics[clip,height=3.0cm]{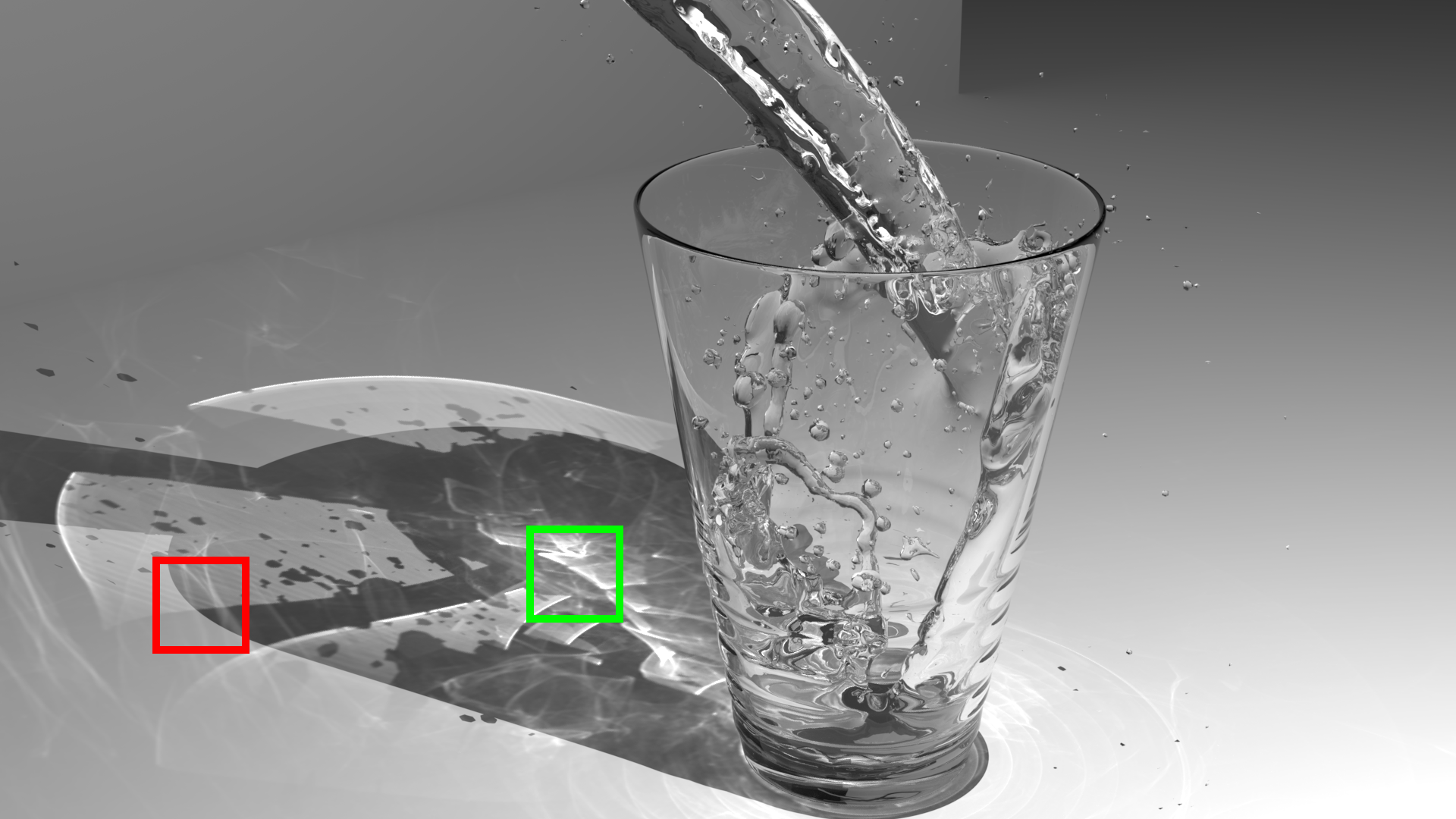}
        \end{minipage}
        &
        \begin{minipage}{0.08\linewidth}
            \includegraphics[height=1.5cm]{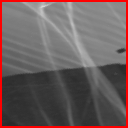}
            \newline
            \includegraphics[height=1.5cm]{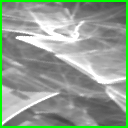}
        \end{minipage}
            &
        \begin{minipage}{0.08\linewidth}
            \includegraphics[height=1.5cm]{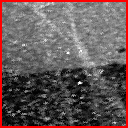}
            \newline
            \includegraphics[height=1.5cm]{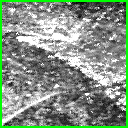}
        \end{minipage}
            &
        \begin{minipage}{0.08\linewidth}
            \includegraphics[height=1.5cm]{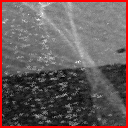}
            \newline
            \includegraphics[height=1.5cm]{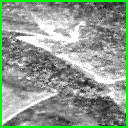}
        \end{minipage}
            &
        \begin{minipage}{0.08\linewidth}
            \includegraphics[height=1.5cm]{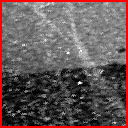}
            \newline
            \includegraphics[height=1.5cm]{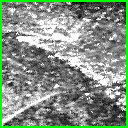}
        \end{minipage}
            &
        \begin{minipage}{0.08\linewidth}
            \includegraphics[height=1.5cm]{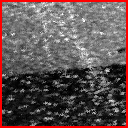}
            \newline
            \includegraphics[height=1.5cm]{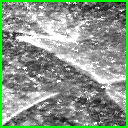}
        \end{minipage}
            &
        \begin{minipage}{0.08\linewidth}
            \includegraphics[height=1.5cm]{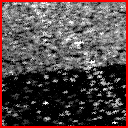}
            \newline
            \includegraphics[height=1.5cm]{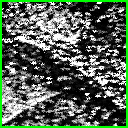}
        \end{minipage}
            \\
            &MSE $\times 10^{2}$:&&0.93&\textbf{0.17}&0.95&1.05&4.32\\
            &(1-SSIM) $\times 10^1$:&&1.94&\textbf{0.81}&1.95&2.22&3.58\\
        \rotatebox[origin=c]{90}{(b) \textsc{Dragon}}&
        \begin{minipage}{0.3\linewidth}
            \centering
            \includegraphics[height=3.0cm]{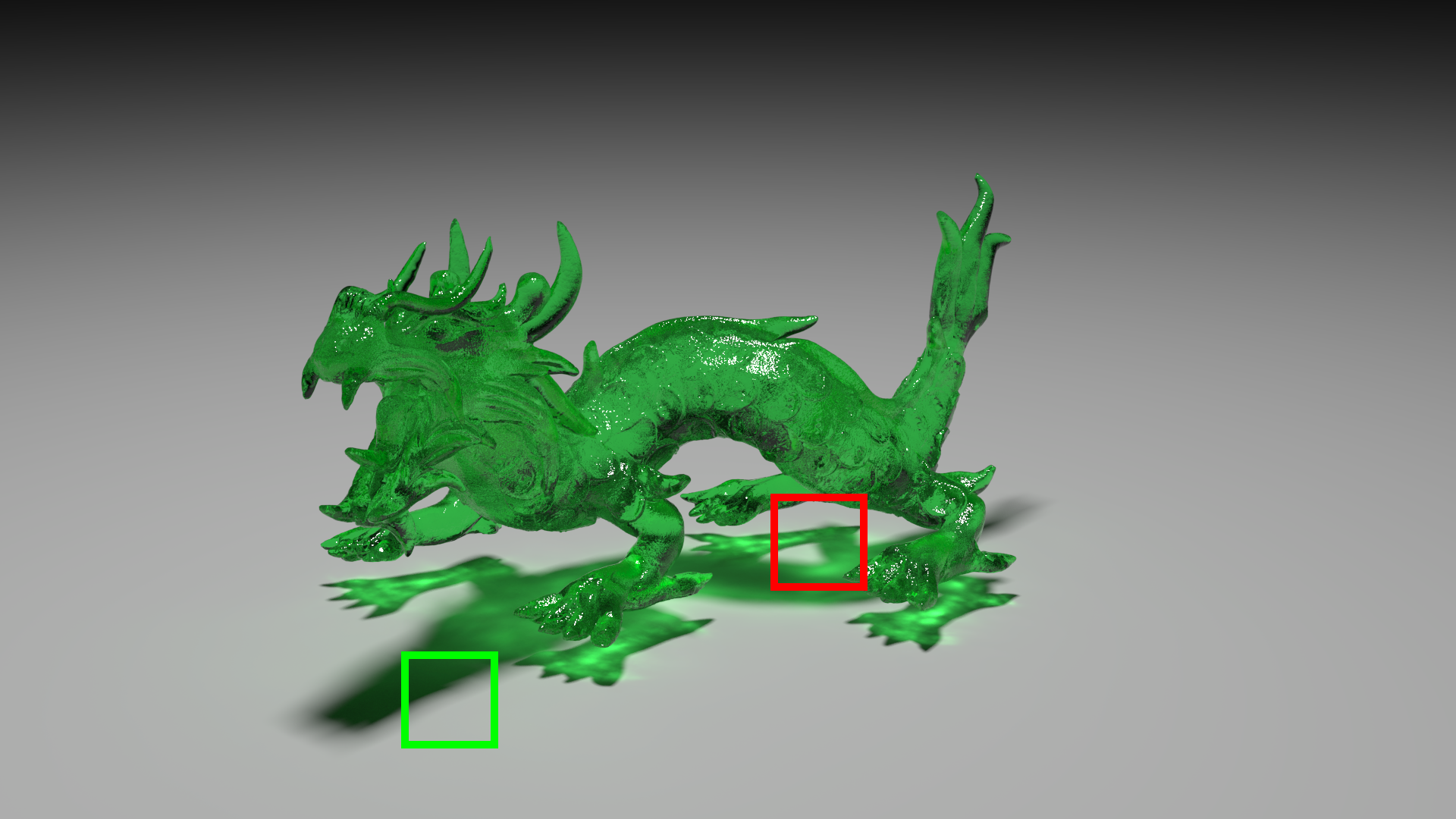}
        \end{minipage}
        &
        \begin{minipage}{0.08\linewidth}
            \includegraphics[height=1.5cm]{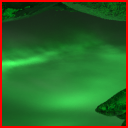}
            \newline
            \includegraphics[height=1.5cm]{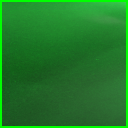}
        \end{minipage}
            &
        \begin{minipage}{0.08\linewidth}
            \includegraphics[height=1.5cm]{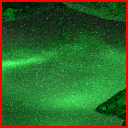}
            \newline
            \includegraphics[height=1.5cm]{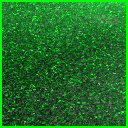}
        \end{minipage}
            &
        \begin{minipage}{0.08\linewidth}
            \includegraphics[height=1.5cm]{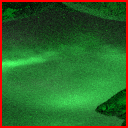}
            \newline
            \includegraphics[height=1.5cm]{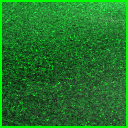}
        \end{minipage}
            &
        \begin{minipage}{0.08\linewidth}
            \includegraphics[height=1.5cm]{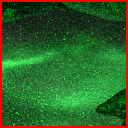}
            \newline
            \includegraphics[height=1.5cm]{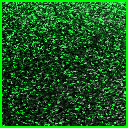}
        \end{minipage}
            &
        \begin{minipage}{0.08\linewidth}
            \includegraphics[height=1.5cm]{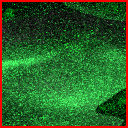}
            \newline
            \includegraphics[height=1.5cm]{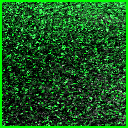}
        \end{minipage}
            &
        \begin{minipage}{0.08\linewidth}
            \includegraphics[height=1.5cm]{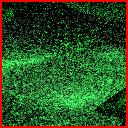}
            \newline
            \includegraphics[height=1.5cm]{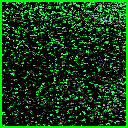}
        \end{minipage}
            \\
            &MSE $\times 10^{3}$:&&0.11&\textbf{0.04}&0.53&0.67&1.46\\
            &(1-SSIM) $\times 10^2$:&&0.87&\textbf{0.31}&2.96&3.43&5.32\\
        \rotatebox[origin=c]{90}{(c) \textsc{BistroStreet}}&
        \begin{minipage}{0.3\linewidth}
            \centering
            \includegraphics[height=3.0cm]{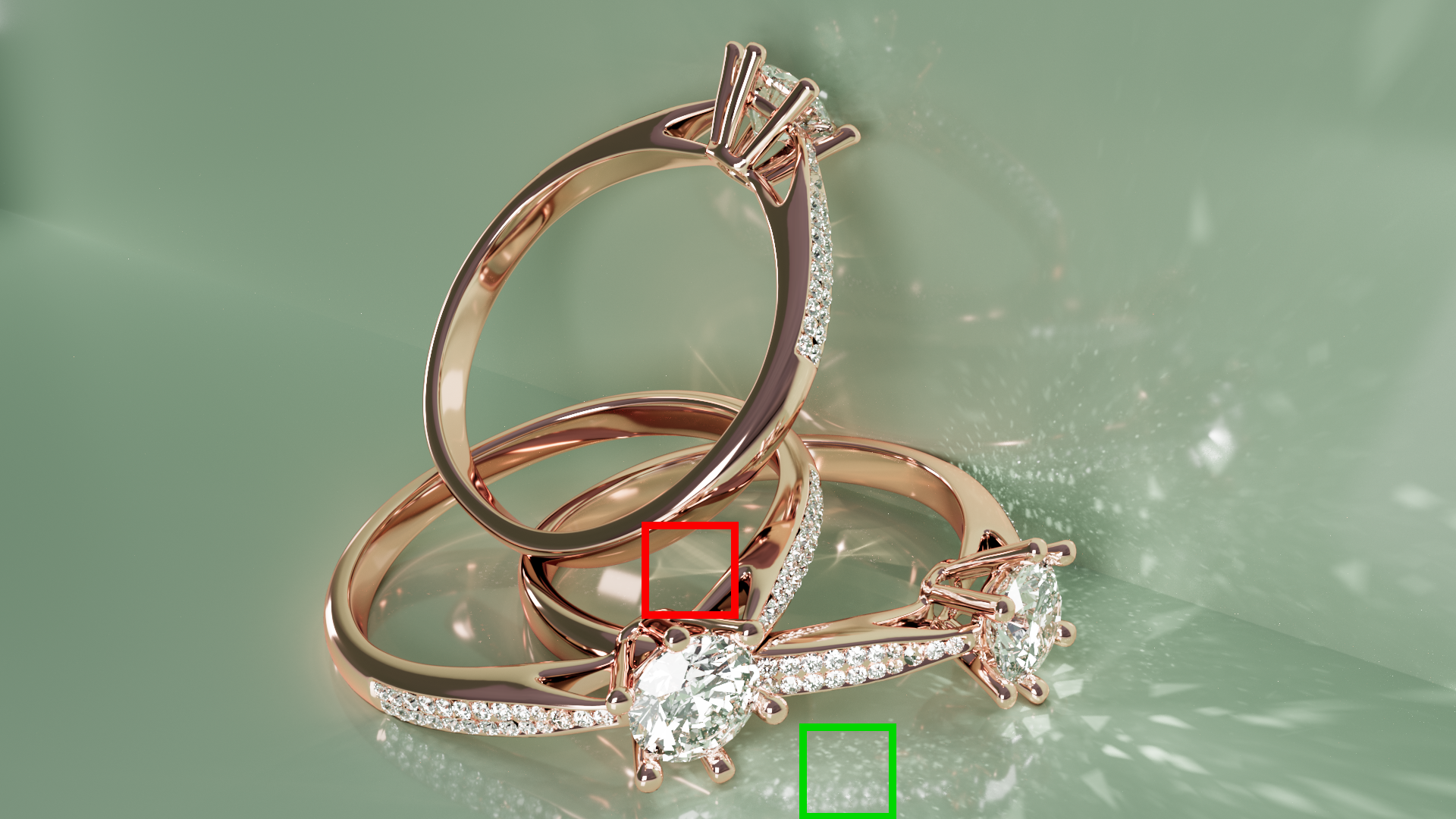}
        \end{minipage}
        &
        \begin{minipage}{0.08\linewidth}
            \includegraphics[height=1.5cm]{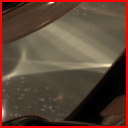}
            \newline
            \includegraphics[height=1.5cm]{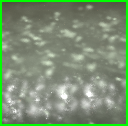}
        \end{minipage}
            &
        \begin{minipage}{0.08\linewidth}
            \includegraphics[height=1.5cm]{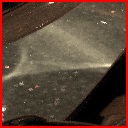}
            \newline
            \includegraphics[height=1.5cm]{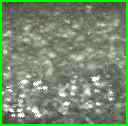}
        \end{minipage}
            &
        \begin{minipage}{0.08\linewidth}
            \includegraphics[height=1.5cm]{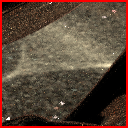}
            \newline
            \includegraphics[height=1.5cm]{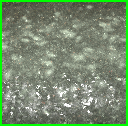}
        \end{minipage}
            &
        \begin{minipage}{0.08\linewidth}
            \includegraphics[height=1.5cm]{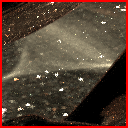}
            \newline
            \includegraphics[height=1.5cm]{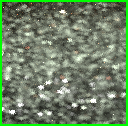}
        \end{minipage}
            &
        \begin{minipage}{0.08\linewidth}
            \includegraphics[height=1.5cm]{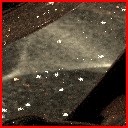}
            \newline
            \includegraphics[height=1.5cm]{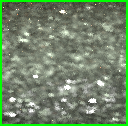}
        \end{minipage}
            &
        \begin{minipage}{0.08\linewidth}
            \includegraphics[height=1.5cm]{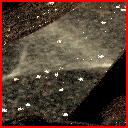}
            \newline
            \includegraphics[height=1.5cm]{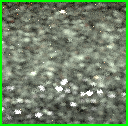}
        \end{minipage}
            \\
& MSE $\times 10^{2}$ && \textbf{0.33} & 0.95 & 1.11 & 1.06 & 2.00 \\
 & (1-SSIM) $\times 10^{1}$ && \textbf{0.78} & 1.38 & 2.22 & 1.95 & 2.23 \\
        \end{tabular}
    \caption{A portion of rendering results in comparison of rendering quality and efficiency across different distributions. The exposure of each inset is adjusted for better observation. (a) Similar results are achieved among G3D, H2D, and vMF under a point light scenario, indicating comparable guiding efficiency with the absence of parallax issues. (b) Under a small rectangular area light, results vary; G3D outperforms other distributions as it can accurately transform global positional distribution into directional distribution at any emission point without parallax issue. (c) When the distribution inside bounds of the caster mesh is not uniform, efficiency of bound-based guiding becomes low, while G3D can be optimized for a close approximation, achieving lower error.}
    \label{fig:dist_compare}
\end{figure*}

We perform rendering for 256 iterations for each method and compare the resulting output to the reference in order to calculate the differences. The metrics derived from these comparisons are presented in \tabref{dist_results}. The results are interpreted in two separate parts as follows:

\paragraph{Compare with directional distributions}
In the \textsc{GlassOfWater} scene, which features only one point light and lacks the parallax issues associated with H2D and vMF, all distributions perform similarly (see \figref{dist_compare} (a)). However, area lights highlight the limitations of H2D and vMF: as demonstrated in the \textsc{Dragon} scene with a small rectangular light overhead, these directional distributions struggle to learn accurate representations for all emission points, rendering the guiding less efficient (see \figref{dist_compare} (b)). In contrast, G3D adapts robustly, transforming spatial distributions to directional distributions accurately at any emission point. H2D and vMF typically assume point light sources and thus do not trivially guide infinite area lights (e.g., environment maps or image-based lighting). G3D easily guides such light sources since its a spatial distribution. In other production-level scenes with complex lighting setups, G3D generally outperforms H2D and vMF, except in the \textsc{Bistro} scene. Here, despite the light sources being delta spotlights, which typically don't present parallax issues, G3D exhibited higher errors due to inadequate initialization. Subsequent experiments with our scene-geometry-based initializer significantly enhanced G3D's performance.
\paragraph{Compare with bound-based guiding}
In this setup, BG outperformed G3D in 6 out of 8 scenes, attributed to G3D's requirement for optimization. BG provides effective guidance from the start, whereas G3D requires a warm-up phase, reducing its initial efficiency. Subsequent experiments demonstrate that our scene-geometry-based initializer substantially accelerates this optimization, allowing G3D to consistently outperform BG. In the 6 scenes where BG was more effective, the distributions within the bounds were somewhat uniform. However, in scenes like \textsc{Ring} and \textsc{SunTemple}, where lighting needs to focus on specific parts of the caster meshes, G3D excelled even without our initializer (see \figref{dist_compare} (c)). This illustrates that while BG achieves full efficiency immediately, its potential is capped due to the uniform assumption of light distribution. The differences between these distributions are further detailed in \secref{learned_result}.

\begin{table}[h]
\centering
\caption{Comparison of guiding with different distribution models, including 3D Gaussians (G3D), 2D histogram (H2D), 3D von Mises-Fisher (vMF), and traditional uniform distribution (Uni). We compare mean square error (MSE, lower is better) and complement of structural similarity index measure (1-SSIM, lower is better). Best results are highlighted in bold. The time cost is also reported for completeness.}
\begin{tabular}{lrrrrrr}
\toprule
Scene && G3D & BG & H2D & vMF & Uni \\
\midrule
\multirow{2}{*}{\makecell{\textsc{Glass}\\\textsc{Of}\\\textsc{Water}}} & MSE $\times 10^{2}$ & 0.93 & \textbf{0.17} & 0.95 & 1.05 & 4.32 \\
 & (1-SSIM) $\times 10^{1}$ & 1.94 & \textbf{0.81} & 1.95 & 2.22 & 3.58 \\
& Time (s) & 17& 16 & 15 & 17 & 5 \\
\midrule
\multirow{2}{*}{\textsc{Dragon}} & MSE $\times 10^{3}$ & 0.11 & \textbf{0.04} & 0.53 & 0.67 & 1.46 \\
 & (1-SSIM) $\times 10^{2}$ & 0.87 & \textbf{0.31} & 2.96 & 3.43 & 5.42 \\
 & Time (s) & 21& 18& 16 & 16 & 8 \\
\midrule
\multirow{2}{*}{\textsc{Bistro}} & MSE $\times 10^{5}$ & 6.75 & 2.41 & \textbf{2.27} & 3.63 & 5.25 \\
 & (1-SSIM) $\times 10^{3}$ & 6.22 & 3.11 & \textbf{3.06} & 4.02 & 5.47 \\
 & Time (s) & 11 & 20 & 12 & 9& 16 \\
\midrule
\multirow{2}{*}{\makecell{\textsc{Bistro}\\\textsc{Street}}} & MSE & \textbf{0.01} & \textbf{0.01} &  N/A &  N/A & 3.30 \\
 & (1-SSIM) $\times 10^{1}$ & 0.08 & \textbf{0.02} & N/A & N/A & 1.32 \\
 & Time (s) & 21& 23& - & - & 16 \\
\midrule
\multirow{2}{*}{\textsc{Eyes}} & MSE $\times 10^{3}$ & 0.23 & 0.03 &  N/A &  N/A & 1.70 \\
 & (1-SSIM) $\times 10^{2}$ & 1.06 & \textbf{0.17} & N/A & N/A & 3.86 \\
 & Time (s) & 31& 32 & - & - & 29 \\
\midrule
\multirow{2}{*}{\textsc{Kitchen}} & MSE $\times 10^{3}$ & 0.09 & \textbf{0.02} & 0.38 & 7.13 & 1.68 \\
 & (1-SSIM) $\times 10^{2}$ & 0.44 & \textbf{0.13} & 1.11 & 4.29 & 2.81 \\
  & Time (s) & 32& 39 & 30 & 28 & 26 \\
\midrule
\multirow{2}{*}{\textsc{Ring}} & MSE $\times 10^{2}$ & \textbf{0.33} & 0.95 & 1.11 & 1.06 & 2.00 \\
 & (1-SSIM) $\times 10^{1}$ & \textbf{0.78} & 1.38 & 2.22 & 1.95 & 2.23 \\
 & Time (s) & 28& 39 & 26 & 27 & 21 \\
\midrule
\multirow{2}{*}{\makecell{\textsc{Sun}\\\textsc{Temple}}} & MSE & \textbf{0.02} & 0.03 & 0.04 & 1.37 & 0.74 \\
 & (1-SSIM) $\times 10^{1}$ & \textbf{2.78} & 3.32 & 2.83 & 5.31 & 4.58 \\
 & Time (s) & 22& 10 & 24 & 25 & 16 \\
\midrule
\bottomrule
\end{tabular}
\label{tab:dist_results}
\end{table}

\subsection{Ablation Studies}
Our method provides a full framework of photon emission guiding. In addition to fitting 3D Gaussians for light sources and sampling from them, we propose a scene-geometry-based initializer and an adaptive light sampler. In this experiment, we evaluate the improvement of rendering quality introduced by each of the components. The baseline is na\"ive guiding with 3D Gaussians, where the Gaussians are na\"ively initialized, $\beta$ linearly increases from 0 to 0.75 in 128 iterations, and light sources are uniformly sampled. We compare the rendering result of the baseline and that with our scene-geometry-based initializer, and the full framework. Additionally, we revisit the comparison with our bound-based guiding (BG), which previously showed strong performance, to underscore the advancements our method offers. The results are shown in \tabref{component_results}.

\begin{figure*}[h]
\setlength{\abovecaptionskip}{2pt}
    \begin{tabular}{crccccc}
        &&Reference&Base&+ Init&Full&BG
                \\
        \rotatebox[origin=c]{90}{(a) \textsc{BistroStreet}}&
        \begin{minipage}{0.36\linewidth}
            \centering
            \includegraphics[height=3.6cm]{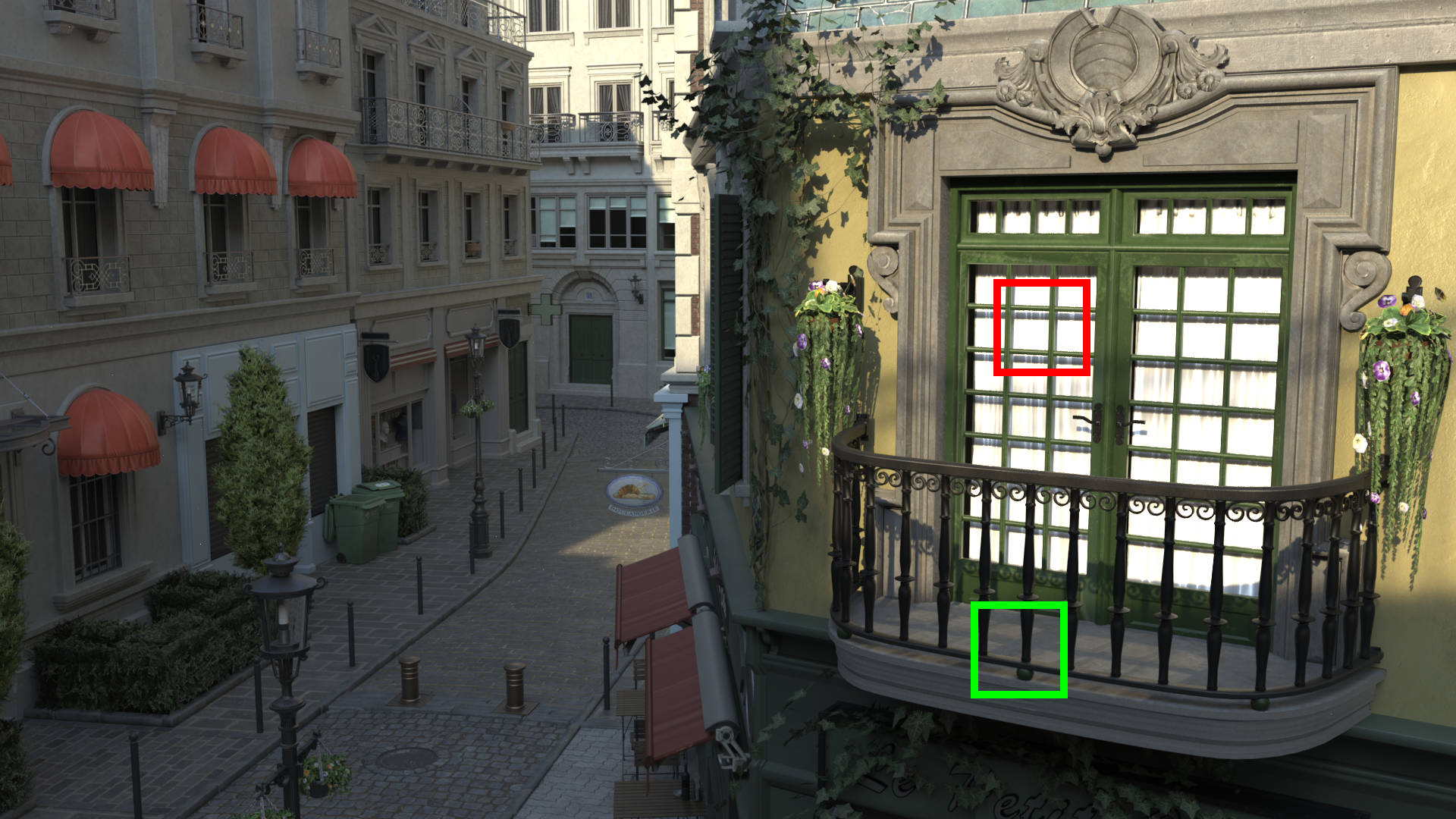}
        \end{minipage}
        &
        \begin{minipage}{0.1\linewidth}
            \includegraphics[height=1.8cm]{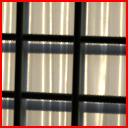}
            \newline
            \includegraphics[height=1.8cm]{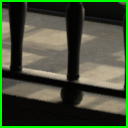}
        \end{minipage}
            &
        \begin{minipage}{0.1\linewidth}
            \includegraphics[height=1.8cm]{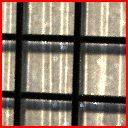}
            \newline
            \includegraphics[height=1.8cm]{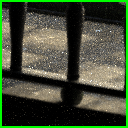}
        \end{minipage}
            &
        \begin{minipage}{0.1\linewidth}
            \includegraphics[height=1.8cm]{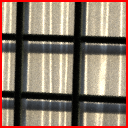}
            \newline
            \includegraphics[height=1.8cm]{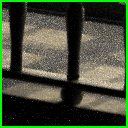}
        \end{minipage}
            &
        \begin{minipage}{0.1\linewidth}
            \includegraphics[height=1.8cm]{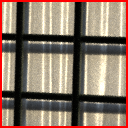}
            \newline
            \includegraphics[height=1.8cm]{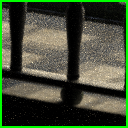}
        \end{minipage}
            &
        \begin{minipage}{0.1\linewidth}
            \includegraphics[height=1.8cm]{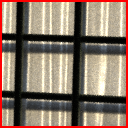}
            \newline
            \includegraphics[height=1.8cm]{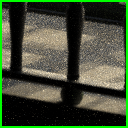}
        \end{minipage}
            \\
& MSE $\times 10^{2}$ && 1.16 & \textbf{0.19} & \textbf{0.19} & 0.24 \\
 & (1-SSIM) $\times 10^{3}$ && 7.76 & \textbf{1.59} & 1.60 & 1.89
        \\
        \rotatebox[origin=c]{90}{(b) \textsc{Eyes}}&
        \begin{minipage}{0.36\linewidth}
            \centering
            \includegraphics[height=3.6cm]{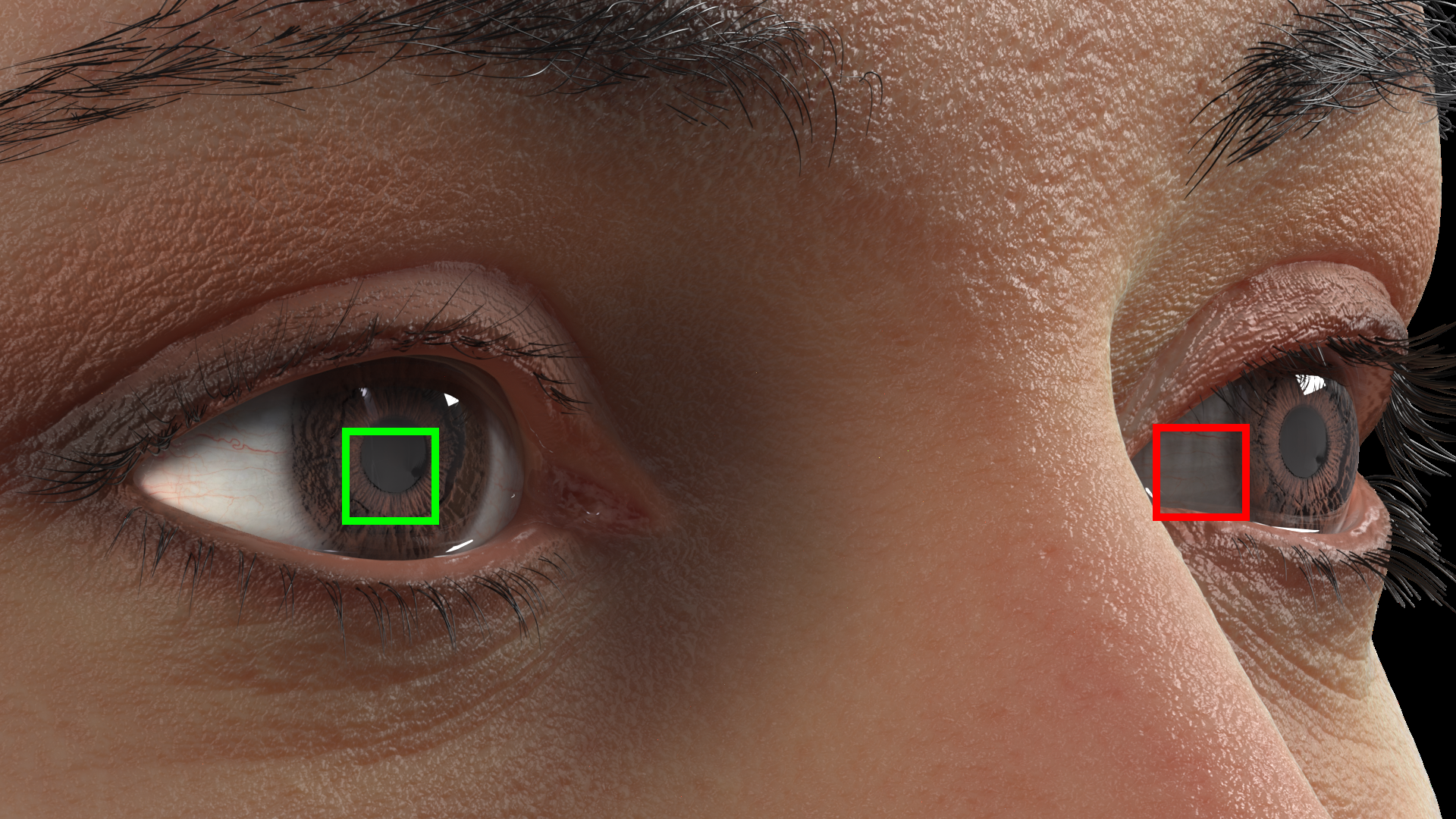}
        \end{minipage}
        &
        \begin{minipage}{0.1\linewidth}
            \includegraphics[height=1.8cm]{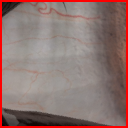}
            \newline
            \includegraphics[height=1.8cm]{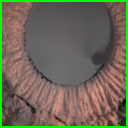}
        \end{minipage}
            &
        \begin{minipage}{0.1\linewidth}
            \includegraphics[height=1.8cm]{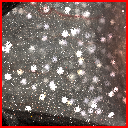}
            \newline
            \includegraphics[height=1.8cm]{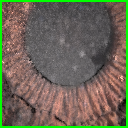}
        \end{minipage}
            &
        \begin{minipage}{0.1\linewidth}
            \includegraphics[height=1.8cm]{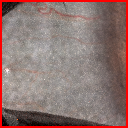}
            \newline
            \includegraphics[height=1.8cm]{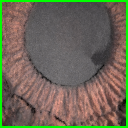}
        \end{minipage}
            &
        \begin{minipage}{0.1\linewidth}
            \includegraphics[height=1.8cm]{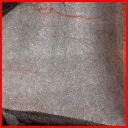}
            \newline
            \includegraphics[height=1.8cm]{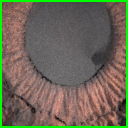}
        \end{minipage}
            &
        \begin{minipage}{0.1\linewidth}
            \includegraphics[height=1.8cm]{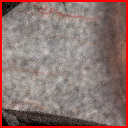}
            \newline
            \includegraphics[height=1.8cm]{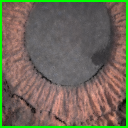}
        \end{minipage}
            \\
            &MSE $\times 10^{4}$:&&2.40&0.17&\textbf{0.14}&0.27\\
            &(1-SSIM)$\times10^{2}$:&&1.10&0.12&\textbf{0.10}&0.17
        \end{tabular}
    \caption{Comparison of rendering results using different components of our photon emission guiding framework. From left to right: reference image, baseline with na\"ive 3D Gaussian guiding, our scene-geometry-based initializer added, the full framework with both the initializer and adaptive light sampler, and the result of bound-based guiding (BG). Exposure of the insets are adjusted for better observation. (a) Our initializer assigns close initial values to the distribution to achieve faster convergence for optimization. With accelerated optimization, G3D outperforms BG. (b) The light sampler in the full framework further enhances the results in rendering tasks with complex lighting setup.}
    \label{fig:ablation_compare}
\end{figure*}

The results, detailed in \tabref{component_results}, demonstrate that each component distinctly improves rendering quality, contributing to a robust photon guiding framework. In the \textsc{Bistro} scene, for instance, our initializer significantly reduces the MSE from $9.96 \times 10^{-5}$ to $1.27 \times 10^{-5}$, marking an 87\% improvement. Furthermore, when compared to the H2D result from the previous experiment ($2.27 \times 10^{-5}$), our full framework achieves a 49\% lower MSE, recorded at $1.16 \times 10^{-5}$. Similar enhancements are observed in the \textsc{Eyes} scene, as illustrated in \figref{ablation_compare}. 
Notably, with the full integration of these components, our method surpasses BG in 7 out of 8 scenes (tied in the \textsc{Kitchen} scene). With improved optimization efficiency, our method is both more performant and more robust compared to BG. 

\begin{table}[h]
\centering
\caption{Comparison of performance using only 3D Gaussians as a baseline (Base), augmented with our scene-geometry-based initializer (+ Init), and employing our full framework (Full), which incorporates our adaptive light sampler. We also compare with bound-based guiding results (BG). The analysis omits time cost but includes it for completeness. The best results are highlighted in bold. Our initializer significantly enhances optimization efficiency and reduces errors across all tested scenes. With improved optimization efficiency, our method consistently surpasses bound-based guiding regardless of geometry complexity. In scenes with complex lighting arrangements, such as \textsc{Eyes} and \textsc{Ring}, our light sampler further improves guiding quality. With these components, our result outperforms BG in 7 scenes out of 8.}

\begin{tabular}{lrrrrr}
\toprule
Scene && Base & + Init & Full & BG \\
\midrule
\multirow{2}{*}{\makecell{\textsc{Glass}\\\textsc{Of}\\\textsc{Water}}} & MSE $\times 10^{3}$ & 7.82 & \textbf{1.23} & 1.29 & 1.73 \\
 & (1-SSIM) $\times 10^{1}$ & 1.76 & \textbf{0.64} & 0.66 & 0.81 \\
  &Time (s) & 17 & 19 & 19& 16\\
\midrule
\multirow{2}{*}{\textsc{Dragon}} & MSE $\times 10^{4}$ & 1.18 & 0.37 & \textbf{0.37} & 0.39 \\
 & (1-SSIM) $\times 10^{3}$ & 9.62 & 3.09 & \textbf{3.06} & 3.14 \\
 &Time (s) & 21 & 21 & 21& 18\\
\midrule
\multirow{2}{*}{\textsc{Bistro}} & MSE $\times 10^{5}$ & 9.96 & 1.27 & \textbf{1.16} & 2.41 \\
 & (1-SSIM) $\times 10^{3}$ & 7.09 & \textbf{1.92} & 1.97 & 3.11 \\
 &Time (s) & 11 & 20 & 21& 20\\
\midrule
\multirow{2}{*}{\makecell{\textsc{Bistro}\\\textsc{Street}}} & MSE $\times 10^{2}$ & 1.16 & \textbf{0.19} & \textbf{0.19} & 0.24 \\
 & (1-SSIM) $\times 10^{3}$ & 7.76 & \textbf{1.59} & 1.60 & 1.89 \\
 &Time (s) & 21 & 29 & 29& 23\\
\midrule
\multirow{2}{*}{\textsc{Eyes}} & MSE $\times 10^{4}$ & 2.40 & 0.17 & \textbf{0.14} & 0.27 \\  
 & (1-SSIM) $\times 10^{2}$ & 1.10 & 0.12 & \textbf{0.10} & 0.17 \\
 &Time (s) & 31 & 32 & 34& 32\\
\midrule
\multirow{2}{*}{\textsc{Kitchen}} & MSE $\times 10^{5}$ & 7.93 & 2.51 & 1.71 & \textbf{1.60} \\
 & (1-SSIM) $\times 10^{3}$ & 3.91 & 1.75 & \textbf{0.94} & 1.35 \\
 &Time (s) & 27 & 36 & 37& 39\\
\midrule
\multirow{2}{*}{\textsc{Ring}} & MSE $\times 10^{3}$ & 3.30 & 8.93 & \textbf{2.63} & 9.52 \\  
 & (1-SSIM) $\times 10^{1}$ & 0.79 & 0.86 & \textbf{0.57} & 1.38 \\
 &Time (s) & 28 & 28 & 31& 39\\
\midrule
\multirow{2}{*}{\makecell{\textsc{Sun}\\\textsc{Temple}}} & MSE $\times 10^{2}$ & 1.89 & 0.85 & \textbf{0.76} & 2.58 \\
 & (1-SSIM) $\times 10^{1}$ & 2.70 & \textbf{1.44} & 1.47 & 3.32 \\
 &Time (s) & 22 & 29 & 29& 10\\
\midrule
\bottomrule
\end{tabular}

\label{tab:component_results}
\end{table}

\subsection{State-of-the-art Comparison}
Two principal methods have been employed to improve the visibility of photons. The first, adaptive importance sampling with 2D histogram, has been explored in research and adopted in production renderers \cite{LWPM, Hyperion, PathOfWater}. The second, Markov-chain Monte Carlo (MCMC), proposed by Hachisuka and Jensen~\shortcite{Hachisuka:MLT}, has also been adopted in production rendering \cite{Corona:Caustics}. In our preliminary studies, MCMC shows clear superiority when compared with H2D. While we've compared with the 2D histogram approach in \secref{distributions}, in this experiment, we compare our method with MCMC. 

Our MCMC implementation closely follows \cite{Hachisuka:MLT}; however, since the original method was designed for progressive photon mapping, we made necessary modifications for adaptation to our rendering scheme. Our target function, $F_i(\mbv)$, is binary: it returns 1 if a photon $\mbv$ successfully reaches a \textit{visible} receiver surface after several specular bounces on caster surfaces, and 0 otherwise. This robust approach marks caustic casters and receivers explicitly, while photon emission and gathering occur in separate passes. Our GPU integration emits $2^{20}$ photons in parallel per iteration, with each thread maintaining its own MCMC chain. During the bootstrap phase, we emit $2^{26}$ photons (64 times the number of MCMC samplers) with random states to comprehensively explore the sample space, forming an initial state pool. Each MCMC sampler then randomly selects a state from this pool as a diverse yet representative starting point. Given the complex scenes and lighting, the bootstrap phase is crucial for ensuring an even initial sample distribution and reduction of correlation, especially when iteration counts are limited. In our GPU implementation, the bootstrap only takes less than 1 second.

Our MCMC chain implements a replica exchange Monte Carlo method, as adapted from \cite{Hachisuka:MLT}. This approach combines uniform sampling and Markov chain strategies within the same framework. Specifically, the chain first proposes a uniform path, emitting a photon $\mbv'$. If $F_i(\mbv')=1$, the photon is recorded and the path accepted; otherwise, a mutation is applied to the last accepted path, emitting a second photon $\mbv''$. This second photon is recorded and the path is accepted if $F_i(\mbv'')=1$. If both proposals fail, the last accepted photon is recorded. We opted not to implement the adaptive mutation size strategy from \cite{Hachisuka:MLT} due to its requirement for per-sample mutation scale updates, which did not integrate well with our GPU-based SPPM scheme. Instead, we explored the mutation algorithm from \cite{PSSMLT} and Gaussian mutations with varying $\sigma$ values, ultimately selecting Gaussian mutation with $\sigma=0.001$ for its lower overall error. The result is reported in \tabref{sota_results}.

\begin{table}[h]
\centering
\caption{Mean square error (MSE, lower is better) and complement of structural similarity index measure (1-SSIM, lower is better) in the comparison between our method (G3D) and the Markov chain Monte Carlo (MCMC) method proposed in \cite{Hachisuka:MLT}. The best metrics are highlighted in bold. The time cost in this experiment is reported for completeness.}
\begin{tabular}{lrrr}
\toprule
Scene && G3D (ours) & MCMC \\
\midrule
\multirow{2}{*}{\makecell{\textsc{Glass}\\\textsc{Of}\\\textsc{Water}}} & MSE $\times 10^{3}$ & 1.29 & \textbf{0.85} \\
 & (1-SSIM) $\times 10^{2}$ & 6.59 & \textbf{4.39} \\
 & Time (s) & 19 & 22 \\
\midrule
\multirow{2}{*}{\textsc{Dragon}} & MSE $\times 10^{5}$ & 3.68 & \textbf{2.37} \\
 & (1-SSIM) $\times 10^{3}$ & 3.06 & \textbf{1.69} \\
 & Time (s) & 21 & 27 \\
\midrule
\multirow{2}{*}{\textsc{Bistro}} & MSE $\times 10^{5}$ & \textbf{1.16} & 2.34 \\
 & (1-SSIM) $\times 10^{3}$ & \textbf{1.97} & 3.25 \\
 & Time (s) & 21 & 29 \\
\midrule
\multirow{2}{*}{\makecell{\textsc{Bistro}\\\textsc{Street}}} & MSE $\times 10^{3}$ & \textbf{1.94} & 3.36 \\
 & (1-SSIM) $\times 10^{3}$ & \textbf{1.60} & 2.71 \\
 & Time (s) & 29 & 33 \\
\midrule
\multirow{2}{*}{\textsc{Eyes}} & MSE $\times 10^{5}$ & \textbf{1.41} & 2.08 \\
 & (1-SSIM) $\times 10^{3}$ & \textbf{0.97} & 1.29 \\
 & Time (s) & 34 & 40 \\
\midrule
\multirow{2}{*}{\textsc{Kitchen}} & MSE $\times 10^{4}$ & \textbf{0.17} & 1.04 \\
 & (1-SSIM) $\times 10^{3}$ & \textbf{0.94} & 4.71 \\
 & Time (s) & 37 & 39 \\
\midrule
\multirow{2}{*}{\textsc{Ring}} & MSE $\times 10^{3}$ & \textbf{2.63} & 2.91 \\
 & (1-SSIM) $\times 10^{2}$ & \textbf{5.70} & 8.17 \\
 & Time (s) & 31 & 33 \\
\midrule
\multirow{2}{*}{\makecell{\textsc{Sun}\\\textsc{Temple}}} & MSE $\times 10^{2}$ & \textbf{0.76} & 8.31 \\
 & (1-SSIM) $\times 10^{1}$ & \textbf{1.47} & 3.19 \\
 & Time (s) & 29 & 30 \\
\midrule
\bottomrule
\end{tabular}
\label{tab:sota_results}
\end{table}

\begin{figure*}[h]
\setlength{\abovecaptionskip}{2pt}
    \begin{tabular}{crccc}
        &&Reference&G3D (Ours)&MCMC
        \\
        \rotatebox[origin=c]{90}{(a) \textsc{Dragon}}&
        \begin{minipage}{0.4\linewidth}
            \centering
            \includegraphics[height=4.0cm]{figures/Compare/Dragon_Dist/ground_truth_beauty_marked}
        \end{minipage}
        &
        \begin{minipage}{0.12\linewidth}
            \includegraphics[height=2.0cm]{figures/Compare/Dragon_Dist/ground_truth_0}
            \newline
            \includegraphics[height=2.0cm]{figures/Compare/Dragon_Dist/ground_truth_1}
        \end{minipage}
            &
        \begin{minipage}{0.12\linewidth}
            \includegraphics[height=2.0cm]{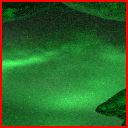}
            \newline
            \includegraphics[height=2.0cm]{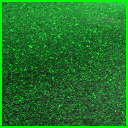}
        \end{minipage}
            &
        \begin{minipage}{0.12\linewidth}
            \includegraphics[height=2.0cm]{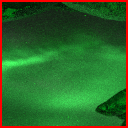}
            \newline
            \includegraphics[height=2.0cm]{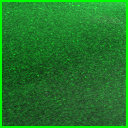}
        \end{minipage}
            \\
            &MSE $\times 10^{5}$:&&3.68&\textbf{2.37}\\
            &(1-SSIM)$\times10^{3}$:&&3.06&\textbf{1.69}
        \\
        \rotatebox[origin=c]{90}{(b) \textsc{Bistro}}&
        \begin{minipage}{0.4\linewidth}
            \centering
            \includegraphics[height=4.0cm]{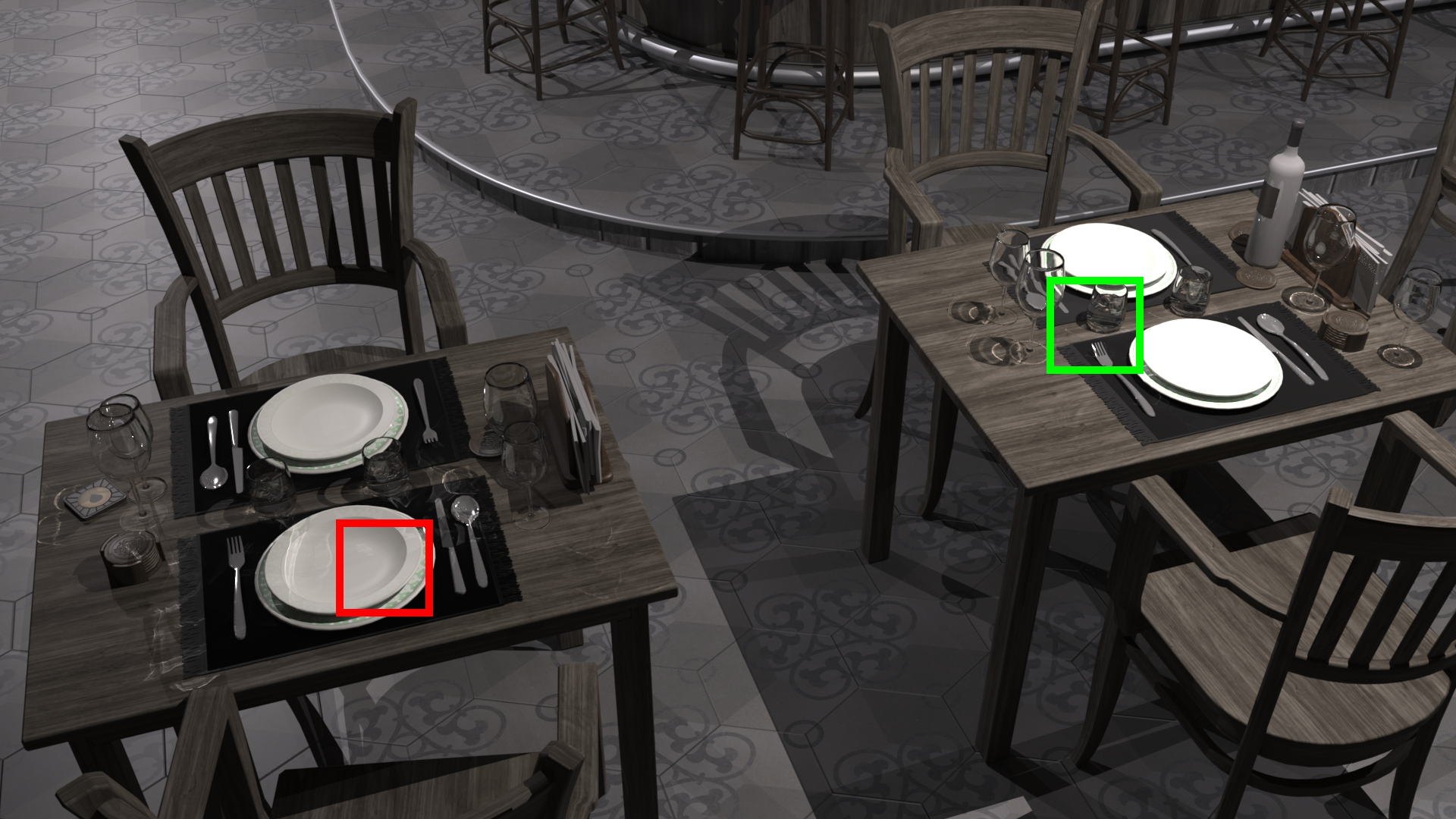}
        \end{minipage}
        &
        \begin{minipage}{0.12\linewidth}
            \includegraphics[height=2.0cm]{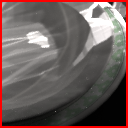}
            \newline
            \includegraphics[height=2.0cm]{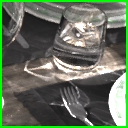}
        \end{minipage}
            &
        \begin{minipage}{0.12\linewidth}
            \includegraphics[height=2.0cm]{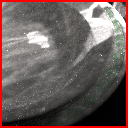}
            \newline
            \includegraphics[height=2.0cm]{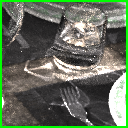}
        \end{minipage}
            &
        \begin{minipage}{0.12\linewidth}
            \includegraphics[height=2.0cm]{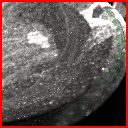}
            \newline
            \includegraphics[height=2.0cm]{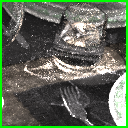}
        \end{minipage}
            \\
            &MSE $\times 10^{5}$:&&\textbf{1.16}&2.34\\
            &(1-SSIM)$\times10^{3}$:&&\textbf{1.97}&3.25
        \\
        \rotatebox[origin=c]{90}{(c) \textsc{SunTemple}}&
        \begin{minipage}{0.4\linewidth}
            \centering
            \includegraphics[height=4.0cm]{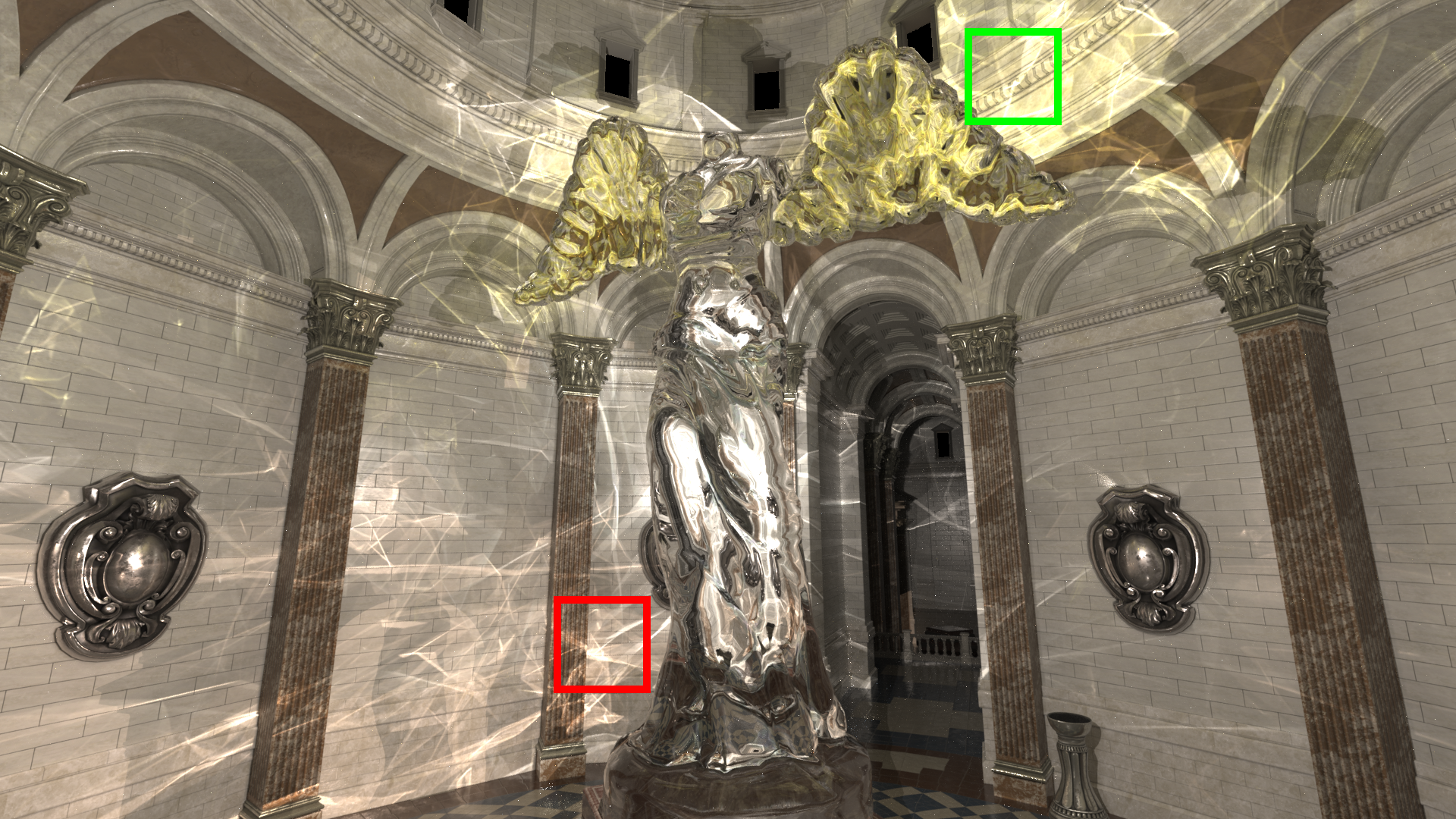}
        \end{minipage}
        &
        \begin{minipage}{0.12\linewidth}
            \includegraphics[height=2.0cm]{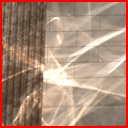}
            \newline
            \includegraphics[height=2.0cm]{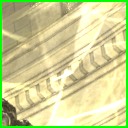}
        \end{minipage}
            &
        \begin{minipage}{0.12\linewidth}
            \includegraphics[height=2.0cm]{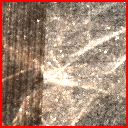}
            \newline
            \includegraphics[height=2.0cm]{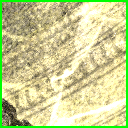}
        \end{minipage}
            &
        \begin{minipage}{0.12\linewidth}
            \includegraphics[height=2.0cm]{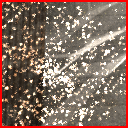}
            \newline
            \includegraphics[height=2.0cm]{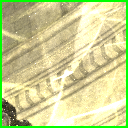}
        \end{minipage}
            \\
            &MSE $\times 10^{2}$:&&\textbf{0.76}&8.31\\
            &(1-SSIM)$\times10^{1}$:&&\textbf{1.47}&3.19
        \end{tabular}
    \caption{Representative results from the comparison between our method (G3D) and the Markov chain Monte Carlo method (MCMC) proposed in \cite{Hachisuka:MLT}. The exposure of each inset is adjusted for better observation. (a) In scenes with straightforward geometry and lighting, MCMC demonstrates higher guiding efficiency than G3D. (b) In a large-scope scene, when only a small portion of the scene casts caustics, MCMC struggles to explore the primary sample space within the given iteration limit. After 256 iterations, the produced result contains slightly higher error than G3D. (c) When a scene contains light sources that have very different visibility distribution, MCMC tends to concentrate on more visible light sources, making it difficult to sample from less visible ones. }
    \label{fig:sota_compare}
\end{figure*}

Readers are referred to the supplemental material for the produced results in full size. In \figref{sota_compare}, we analyze several results that demonstrate the difference between our method and MCMC. 
In scenes with simple geometry and lighting, MCMC shows clear superiority: in \textsc{WaterOfGlass} and \textsc{Dragon}, MCMC achieves lower error and smooth caustics (see \figref{sota_compare} (a)).
However, in complex production-level scenes, MCMC sometimes struggles to effectively explore the primary sample space, leading to slightly higher image errors  (see \figref{sota_compare} (b)). \v{S}ik \etal\shortcite{MCMC:Survey} reported the non-uniform convergence of MCMC methods in complex scenarios, which is consistent with our observation. In rendering the \textsc{SunTemple} scene, as shown in \figref{sota_compare} (c), MCMC encounters inherent limitations that affect accuracy. This scene features two common delta light sources: an infinite directional light and a point light in front of the glass statue. Due to the large size of the scene, the overall visibility of the directional light is approximately 0.002\%, compared to 5.9\% for the point light (2950$\times$ higher). When a uniform light sampler is employed to select light sources, MCMC tends to get stuck at the point light, struggling to adequately explore paths from the directional light. Efforts to address this by adjusting the weights assigned to different light sources proved insufficient. For instance, assigning a weight of 0.99 to the directional light still resulted in noisy outcomes. Increasing the weight to 0.999 allowed for normal sampling of the directional light but introduced a visual bias in the contribution from the point light. The spatial target function proposed by Gruson \etal\shortcite{SpatialTargetFunction} is not effective in this scenario, as the challenge here stems from differences in visibility that are not location-dependent. Thus, while MCMC is highly efficient, it can encounter challenges in scenarios with highly disparate light source visibilities. Our method in contrast performs more robustly and provides a good option for photon mapping in production rendering.

\subsection{Learned Result}\label{sec:learned_result}
\begin{figure*}[h]
\begin{tabular}{cccc}
\includegraphics[width=0.24\linewidth]{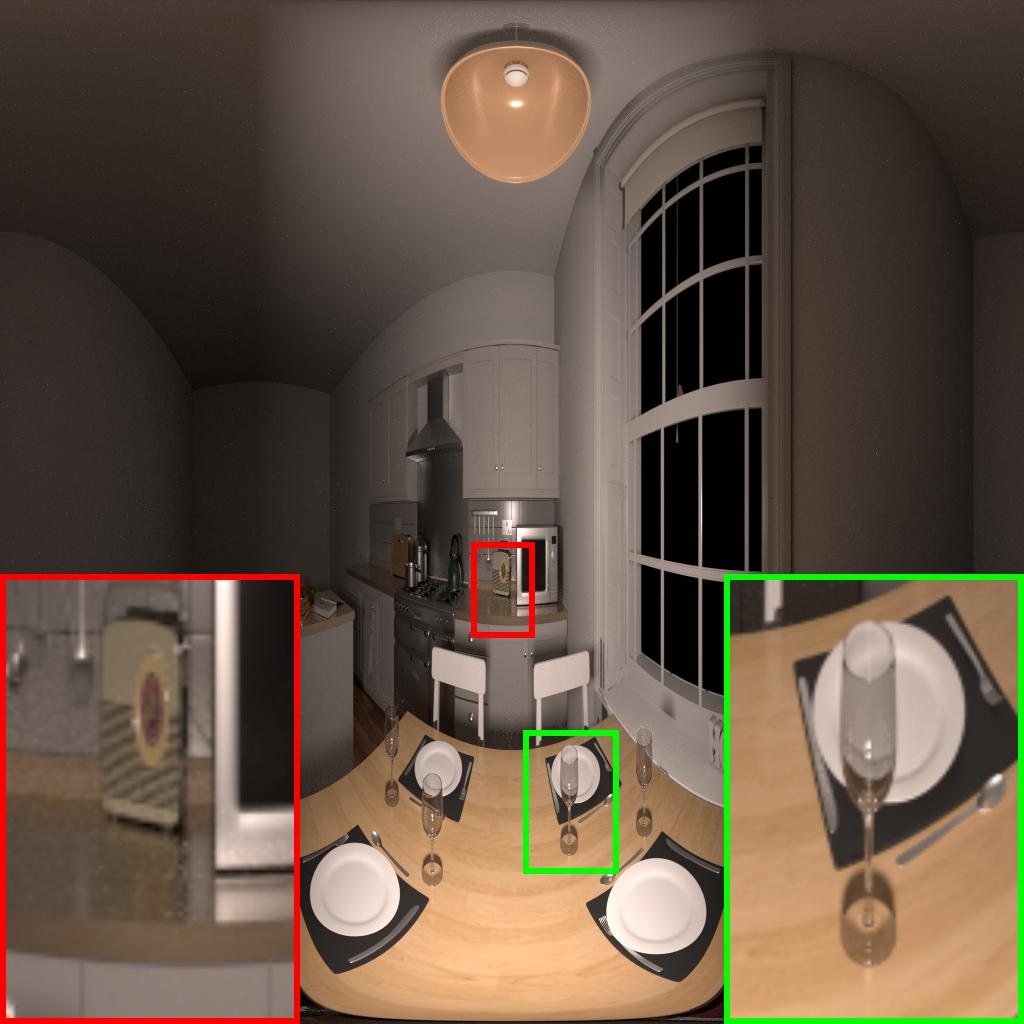}
&
\includegraphics[width=0.24\linewidth]{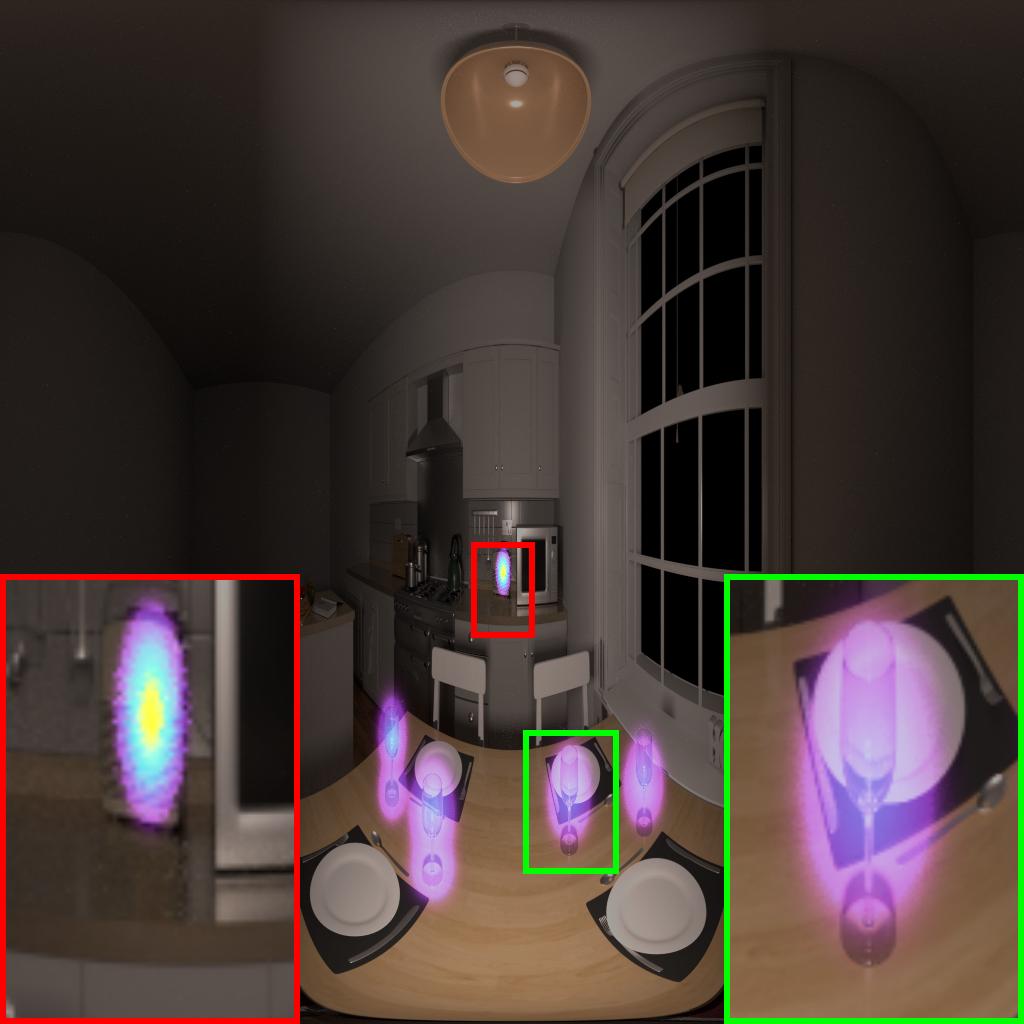}
&
\includegraphics[width=0.24\linewidth]{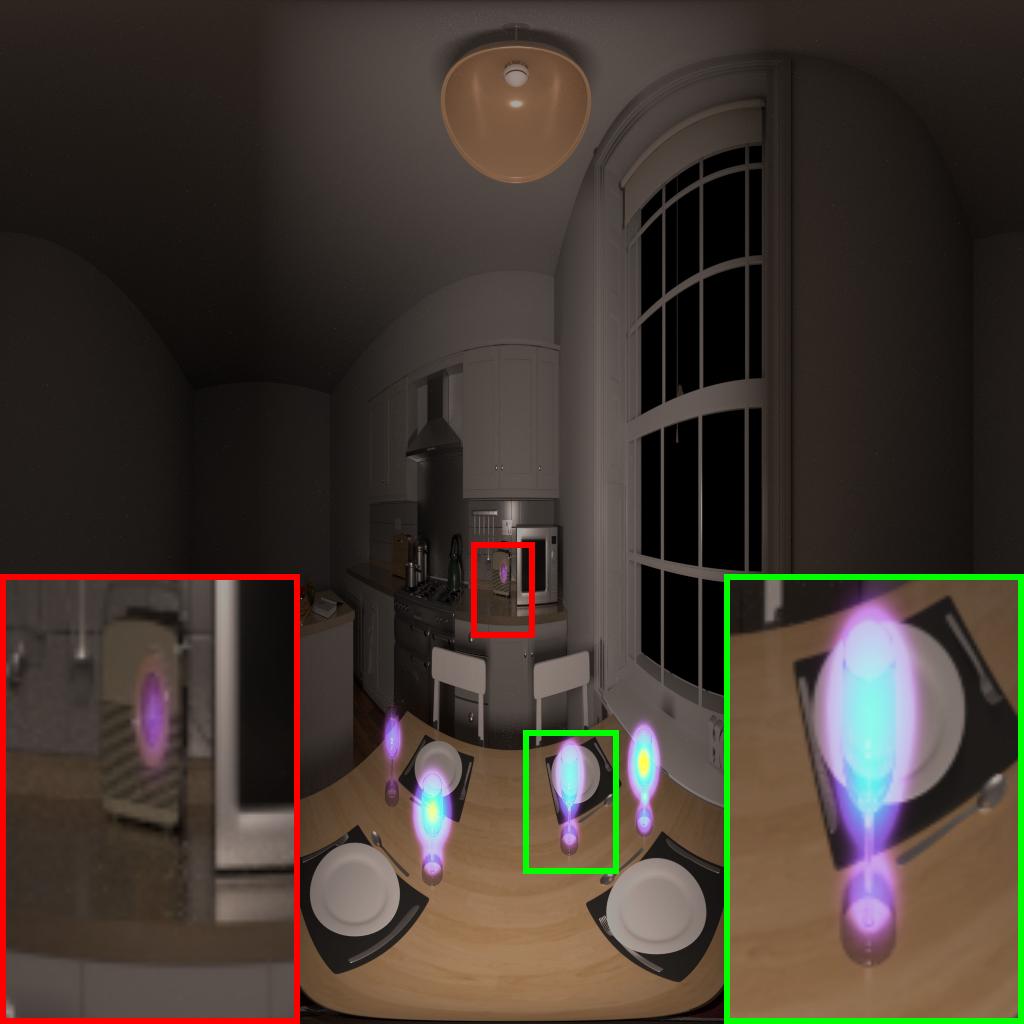}
&
\includegraphics[width=0.24\linewidth]{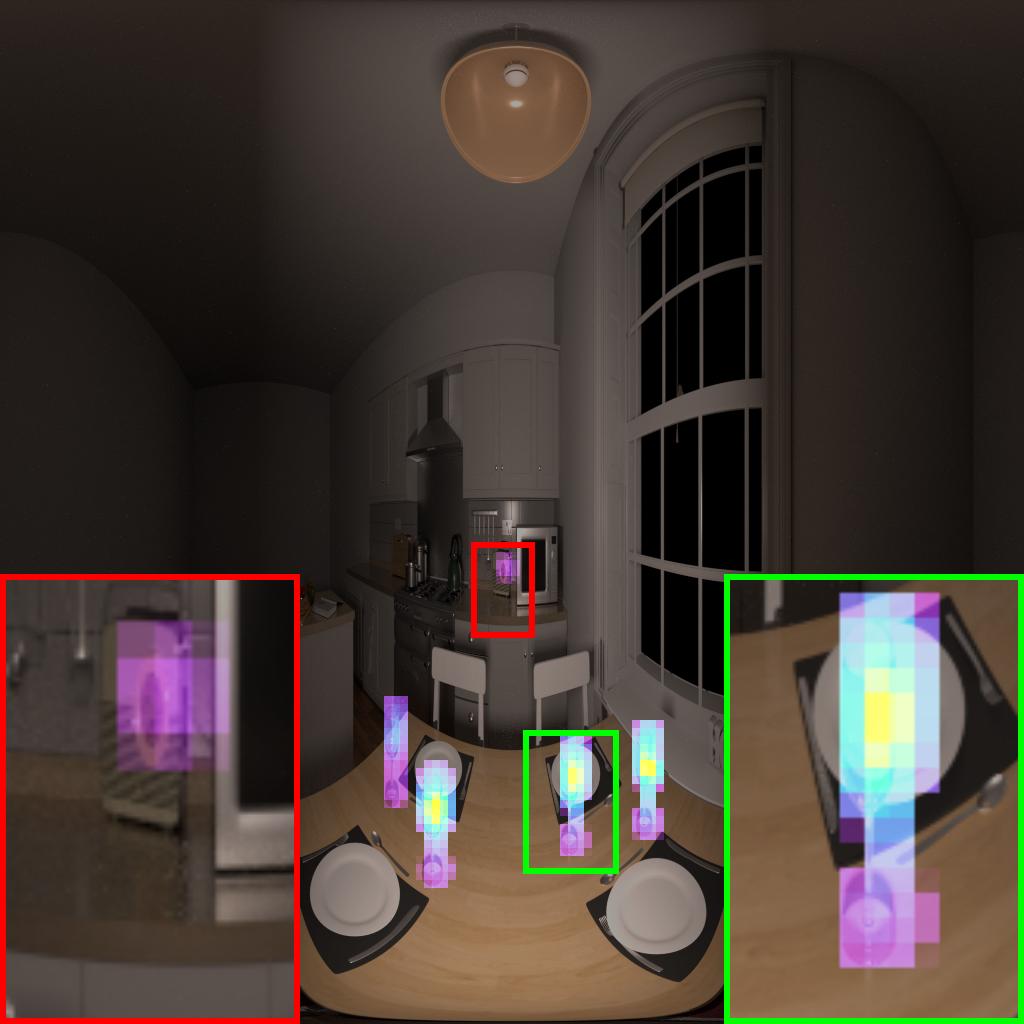}
    \\
    (a)&(b) G3D (ours)&(c) vMF&(d) H2D\\
\includegraphics[width=0.24\linewidth]{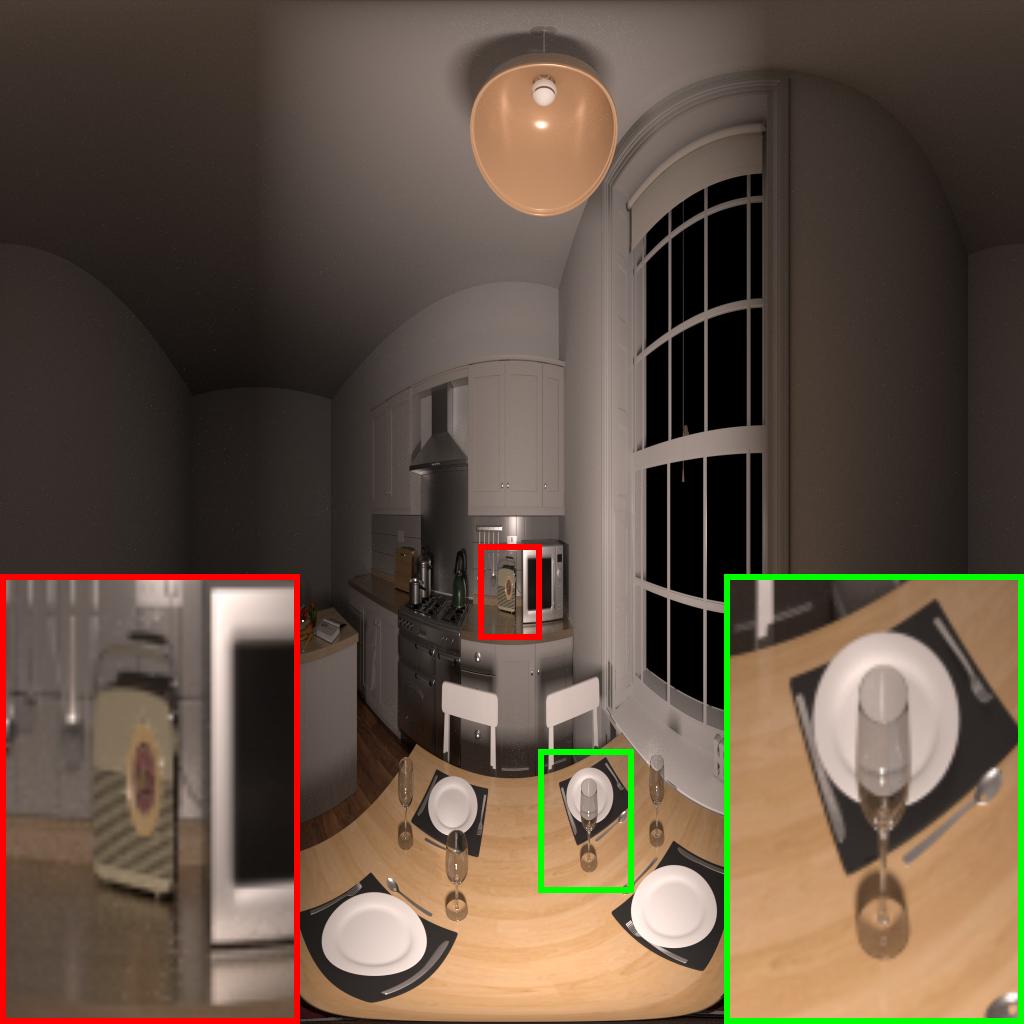}
&
\includegraphics[width=0.24\linewidth]{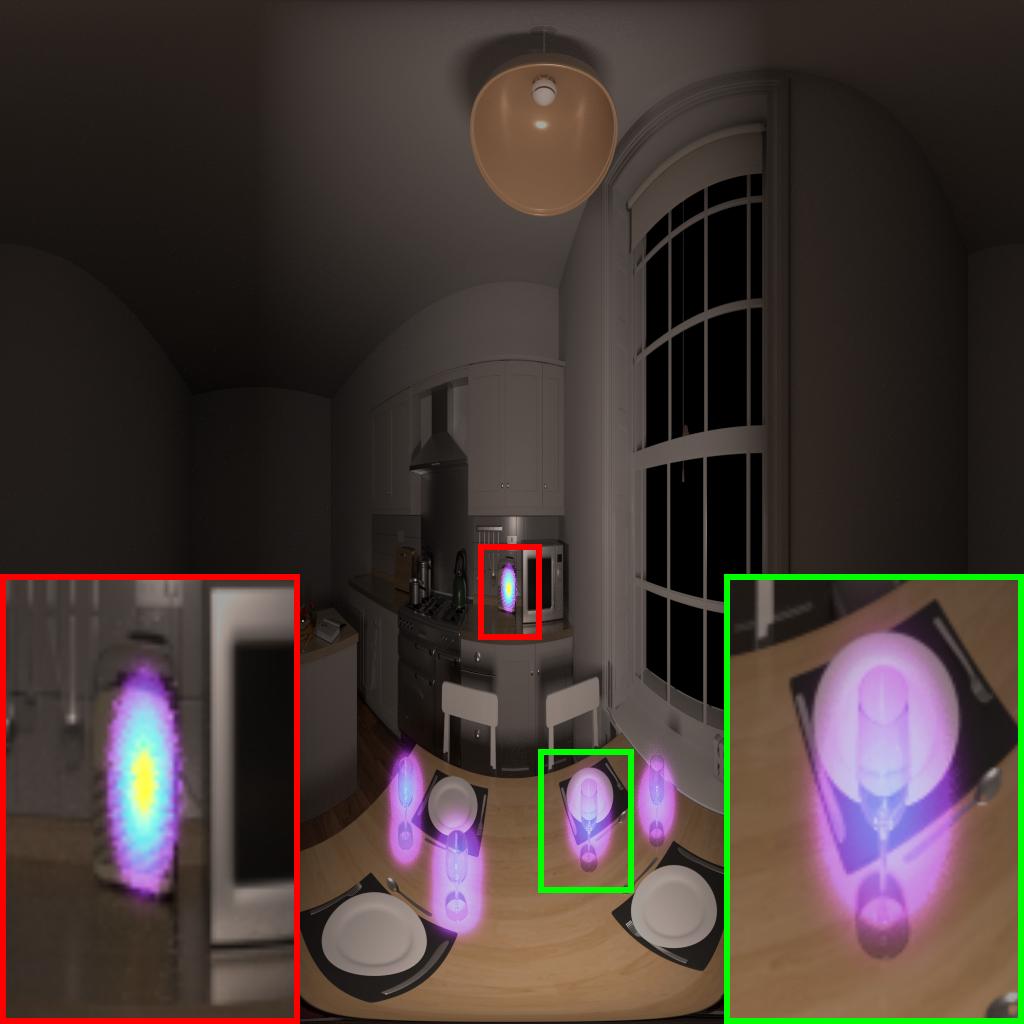}
&
\includegraphics[width=0.24\linewidth]{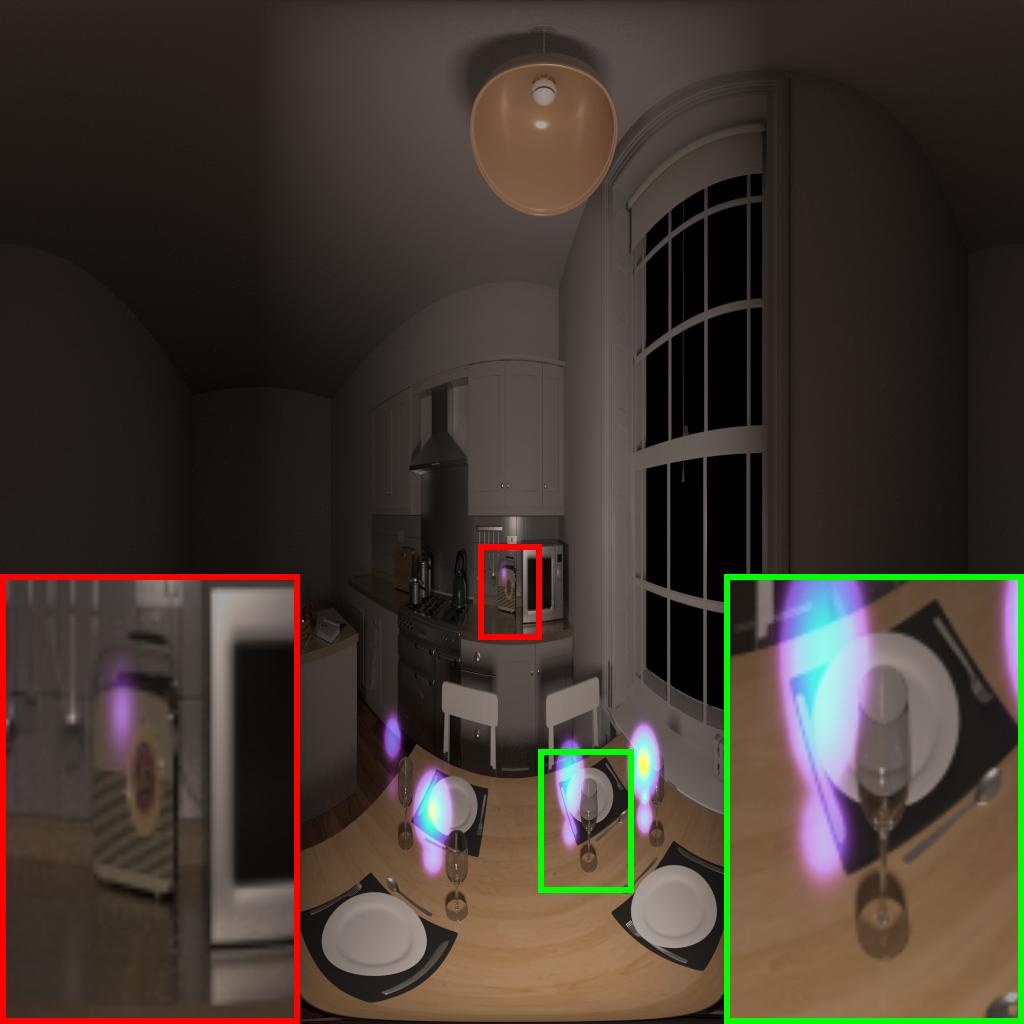}
&
\includegraphics[width=0.24\linewidth]{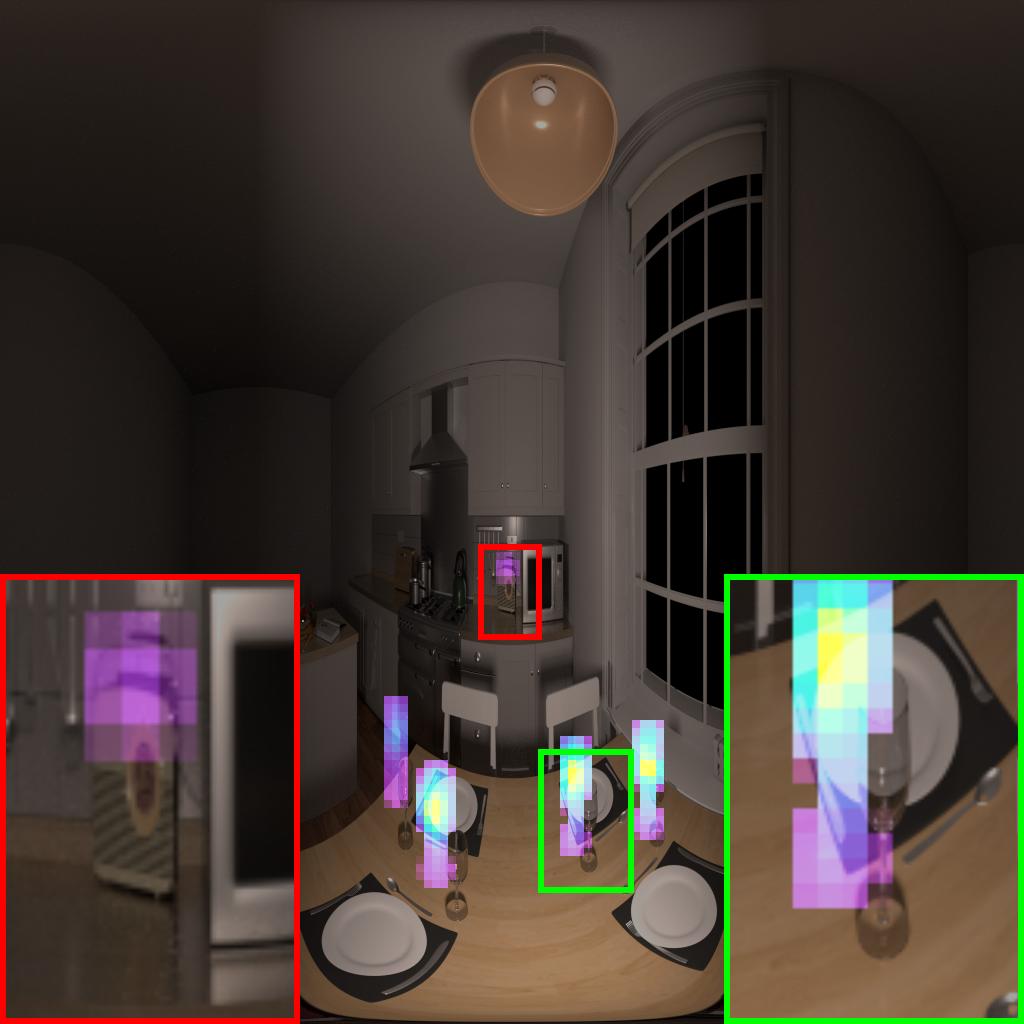}
    \\
    (e) &(f) G3D (ours)&(g) vMF&(h) H2D
\end{tabular}\\
  
    \caption{For a small rectangular light source colocated at the camera position in (a), we fit the emission distribution using G3D, H2D, and vMF models. The resulting emission directions are visualized in (b), (c), and (d), respectively. When emitting from the center of the area light, all models fit the caustics casters accurately (indicated by arrows). However, when emitting from a corner of it, as shown in (e) (note the change in perspective from (a)), the directional models H2D and vMF suffer from parallax issues. This causes their distributions to misalign with the casters, as observed in (g) and (h) respectively (observe the misalignment with the glasses). The G3D model, shown in (f), maintains an accurate distribution because it derives the directional distribution from the global positional distribution. Readers are encouraged to zoom in on the visualizations for a detailed view.}
    \label{fig:parallax_demo}
\end{figure*}

In addition to the quantitative experiments, we conducted an experiment to directly compare the learned distributions using different models. In \figref{parallax_demo}, we demonstrate the parallax issue and how G3D eliminates it. For a small rectangular light source located at the camera position in \figref{parallax_demo} (a) and facing forward, we fit the emission distribution using three models: G3D, H2D, and vMF. When emitting from the center of the rectangular light, as shown in \figref{parallax_demo} (a), the directional distribution of all models accurately fits the actual caster geometry. However, when emitting from a corner of the light, as shown in \figref{parallax_demo} (e), the distributions of vMF and H2D deviate from the actual situation due to the parallax issue. Our method transforms G3D from a spatial distribution into a directional one at any observation point, allowing it to guide emission accurately even from offset positions. Furthermore, this advantage extends to infinite directional lights with directional areas, where similar parallax challenges can occur. For infinite lights, our G3D method generates and samples a 2D distribution corresponding to the sampled direction.

\begin{figure}
\begin{tabular}{cc}
\includegraphics[width=0.45\linewidth]{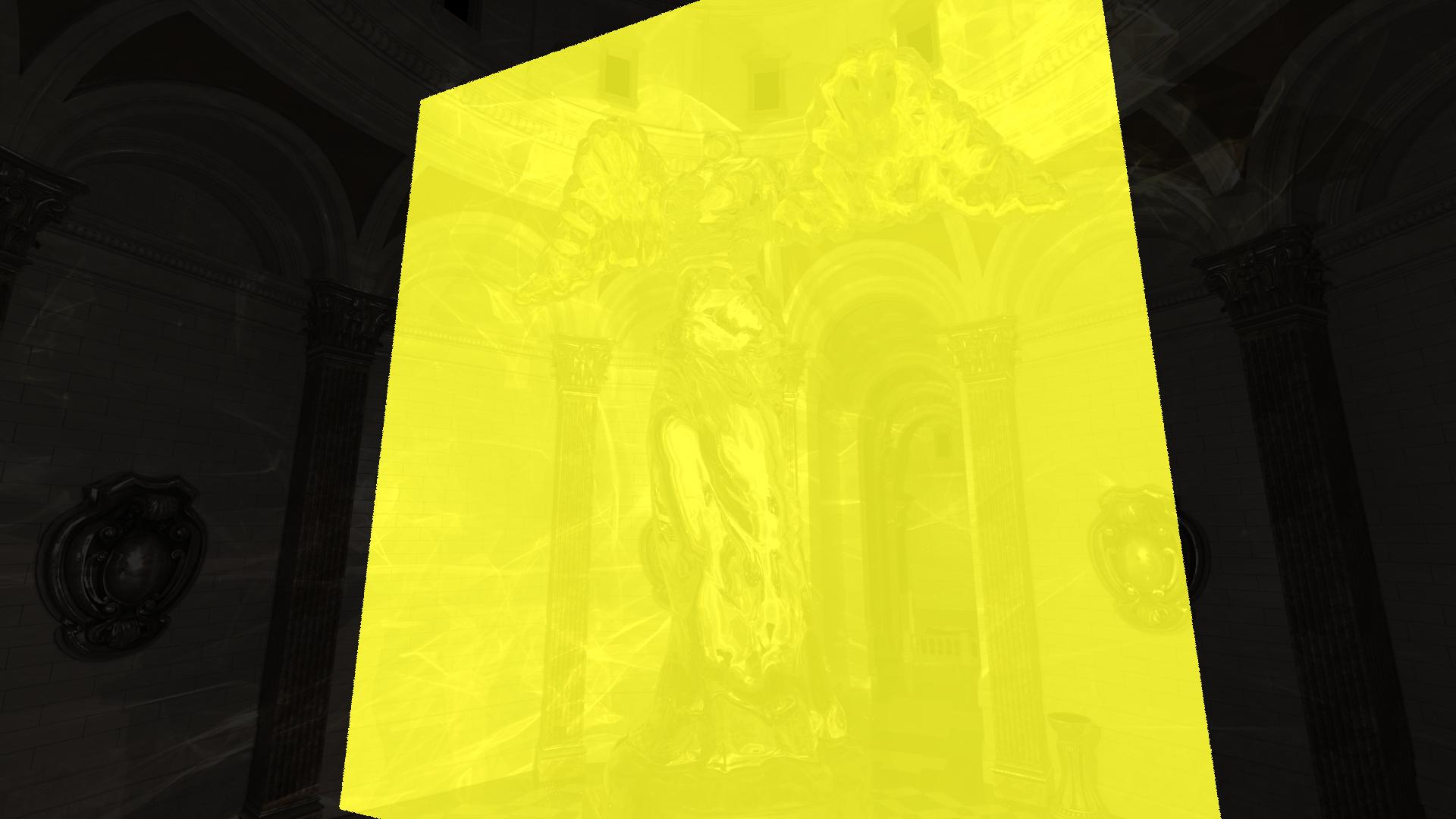}
&
\includegraphics[width=0.45\linewidth]{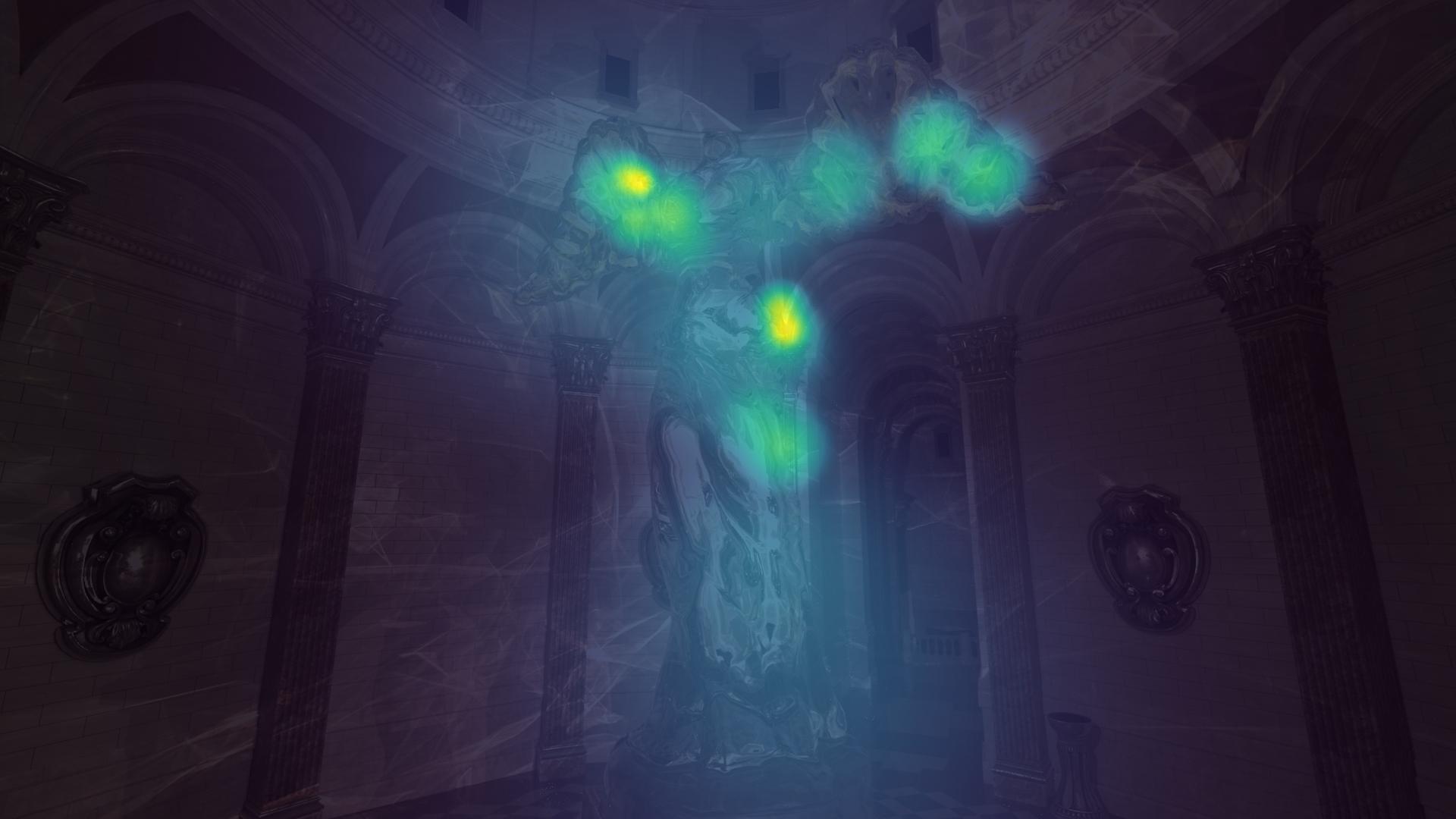}\\
\includegraphics[width=0.45\linewidth]{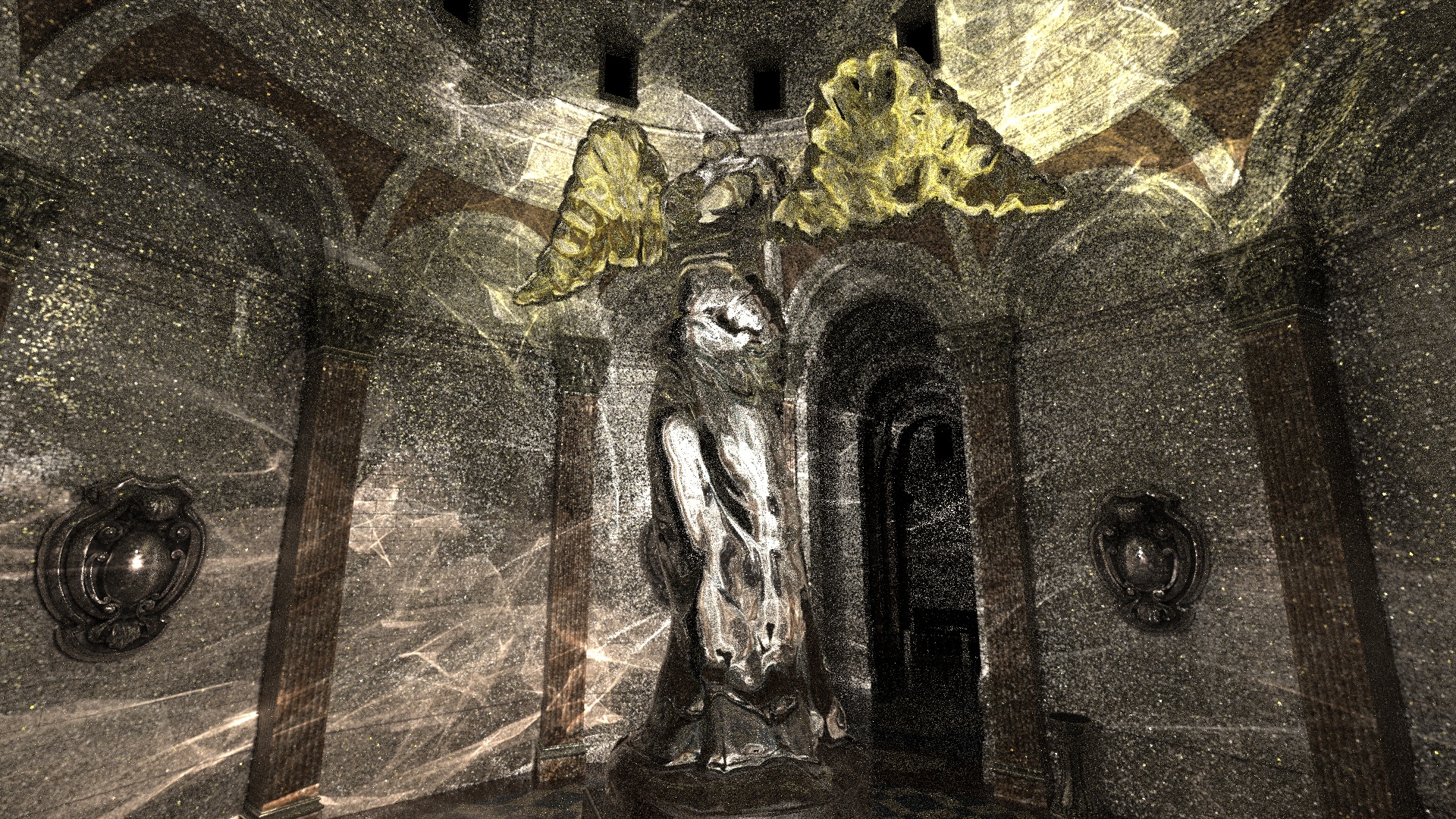}
&
\includegraphics[width=0.45\linewidth]{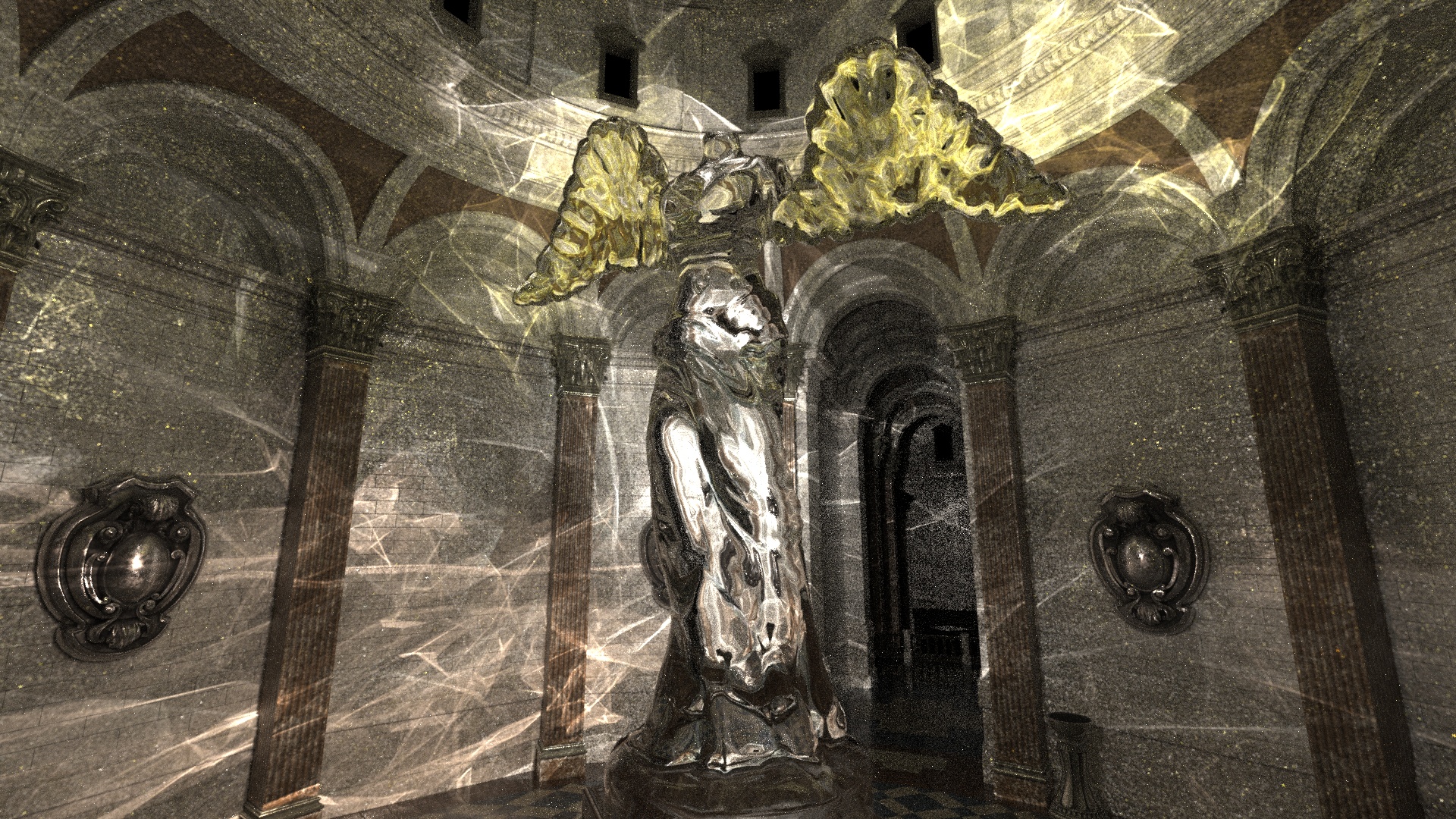}\\
Bound-based guiding&Ours
\end{tabular}\\
\caption{Visualization of fitted 3D Gaussian mixture from all light sources, and the guiding bound used in our bound-based guiding, in the \textsc{SunTemple} scene. Compared to the na\"ive bound-based guiding (left), our 3D Gaussian method effectively focuses on the important part of a single mesh (right), which makes it efficient when only part of the caster geometry needs to be considered, or when the geometry contains a lot of empty space. Readers are encouraged to zoom in to compare the quality of rendering results.}
\label{fig:positional_dist}
\end{figure}

We also visualize the spatial distribution of the learned 3D Gaussian mixtures in \figref{positional_dist}, in comparison with bound-based guiding. The learned distribution demonstrates superiority over bound-based guiding. In the \textsc{SunTemple} scene, the statue geometry contains a lot of empty space within its bounds, and not all of the mesh requires equal even photon emission. Our 3D Gaussian mixture quickly prunes to only the relevant regions to efficiently guide photon emission.

\begin{figure}[h]
\includegraphics[width=0.9\linewidth]{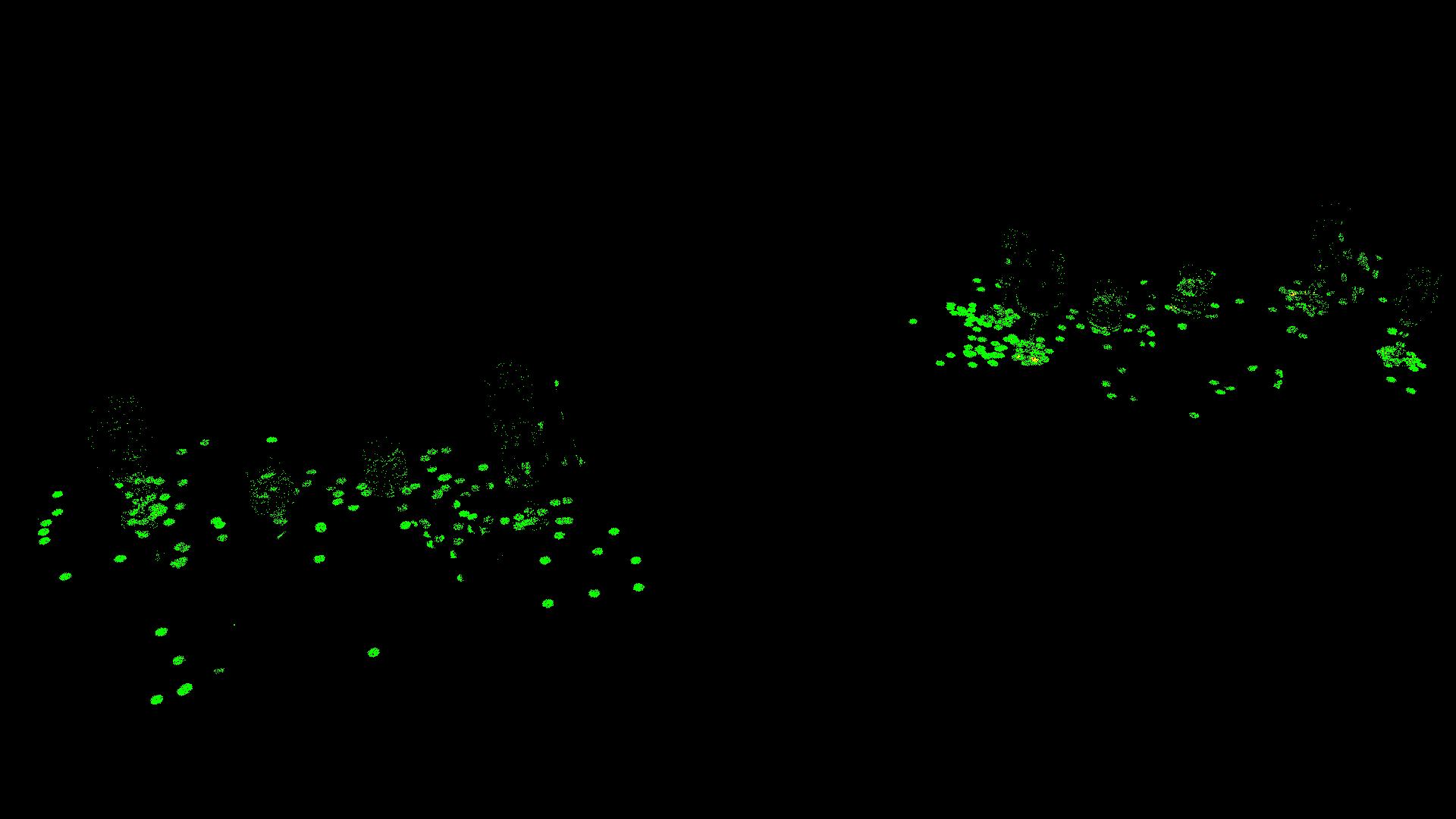}\\
\includegraphics[width=0.9\linewidth]{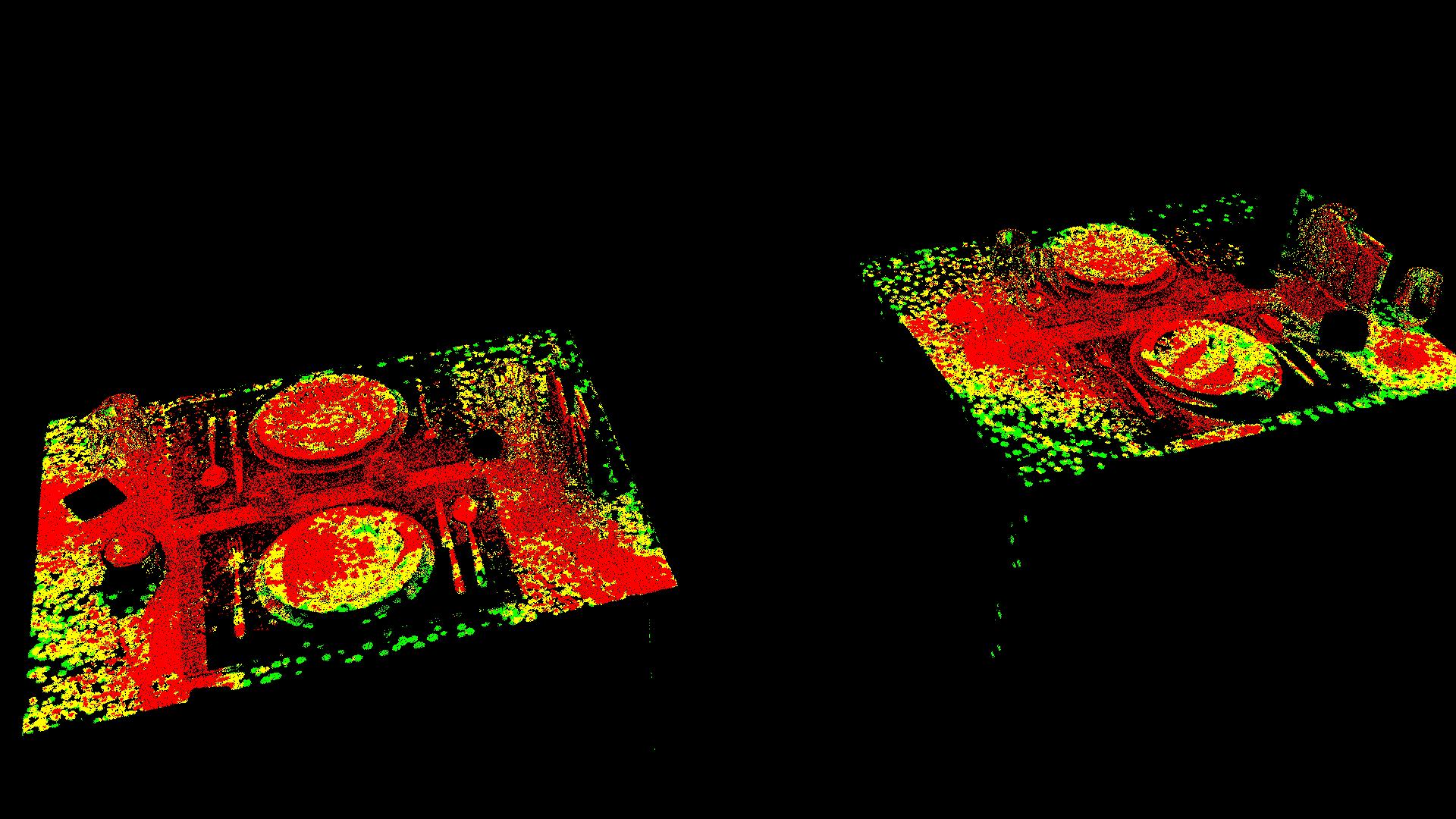}
  
    \caption{Visualization of the searching radius of each pixel that successfully gathers photons during one iteration in the \textsc{Bistro} scene. Green indicates a larger searching radius at the surface, and red indicates a smaller one (hence a higher photon density). With uniform emission (top), only a few photons successfully reach the visible receiver surface, resulting in low or zero photon density for most pixels. Our method (bottom) effectively guides the photons to emit more into the visible region, achieving a more accurate density estimation.}
    \label{fig:photon_vis}
\end{figure}

Finally, in \figref{photon_vis}, we visualize the photon density from one rendering iteration of the same scene. In one rendering iteration of the \textsc{Bistro} scene, our photon guiding method increases the photon count in the visible area from 438 to 30,1326: $687\times$ more visible photons, leading to a significant boost in density and more accurate estimation.

\section{Discussion}
\paragraph{Cosine-weighted guiding}
Theoretically, our method can be extended for cosine-weighted guiding for emission direction. We have shown that the spherical distribution \eqref{directional}, obtained using our novel directional transform of 3D Gaussian, has a similar shape with vMF distribution $F_v$ (while they are essentially different). For an area light, if we can first approximate $F_o$ with $F_v$, then, with another $F_v$ resembling the cosine-weighted distribution of directions with respect to the surface normal, we are able to calculate the product of these two, which is still a vMF distribution. By sampling the product vMF we can achieve cosine-weighted guiding for diffuse area lights. However, finding a $F_v$ approximating $F_o$ is expensive, since it requires solving a transcendental equation that connects parameters of $F_o(\omega~|~d,\sigma)$ and $F_v(\omega~|~\kappa)$. We leave this as a future work. 

\paragraph{4D guiding}
The photon emission can be considered as a 4D distribution, while our method focus on half of the dimensions which we consider is dominating, and leave the rest uniformly sampled. For most actual rendering tasks this is efficient, however, it could suffer when the other half has an extreme distribution. This limitation could be addressed by incorporating another importance sampling technique, for example, it's possible to fit another distribution for position sampling in area lights. We leave this as a future work.

\paragraph{Indirect caustics}
Our method does not handle caustics from indirect lighting. However, it is also possible to extend our method and combine it with path guiding techniques to achieve a solution to this challenge. In such a solution, 3D Gaussians are still leveraged to guide the initial photon emission, and the path guiding technique is used to guide the bounce at the diffuse surfaces. We leave this as a future work too.

\paragraph{Multi-lobe guiding}
In caustics rendering, many photon paths involve multi-lobe BSDF sampling (e.g., glass has both a reflection lobe and a refraction lobe). For such paths, we currently sample lobes with their corresponding BSDF weights, however, the variance could be further reduced if we are able to sample the decision based on the lobes' actual contribution to the image. In fact, in the comparison of \textsc{Pool} scene (\figref{sota_compare} (c)), MLT achieves the best result mainly because it invests more samples on the reflection of the water surface. Adapting a variance-aware distribution, as suggested by Rath et al.~\shortcite{VAPG}, is a conceivable solution. However, the practical implementation of such a concept within our gradient-based learning remains unexplored, which we leave as a future work. 
\section{Conclusion and Future Work}
In this paper, we have introduced a novel photon guiding method designed for production rendering systems, focusing on efficiently rendering caustics. The cornerstone of our approach is the innovative utilization of 3D Gaussian mixture. We have derived a novel transform that enables sampling directions from 3D Gaussians at any observation point, effectively eliminating the common parallax issue associated with existing directional distributions. Additionally, we propose a scene-geometry-based initializer and an adaptive light source sampler to further improve the efficiency of online learning and guiding. Through experiments, we demonstrate that our framework not only outperforms existing directional distributions but is also more robust than state-of-the-art methods in handling complex rendering tasks.

We consider 3D Gaussians to not only serve as the foundation of our method, but also to hold immense potential for a wide range of applications in computer graphics research. For example, the versatility of 3D Gaussians in connecting local and global distributions is also promising in a broader path guiding context, such as modeling indirect lighting and view-dependent visual effects. We will explore such directions in our future work. 

\bibliographystyle{ACM-Reference-Format}
\bibliography{photon_guiding}
\clearpage
\onecolumn
\appendix
\section{Derivation of directional transform}
Consider an unnormalized isotropic 3D Gaussian distribution of the following form with mean vector $\mu$:
\begin{equation}\label{eq:3DGaussian}
	G(\mathbf{x})=\exp\!\left(-\frac{\|\mathbf{x}-\mu\|^2}{2\sigma^2}\right).
\end{equation}
Let $\mathbf{x}_0$ be the location of observation point.
Set
\begin{equation*}
	d=\|\mathbf{x}_0-\mu\|.
\end{equation*}
We consider a polar coordinate centered at $\mathbf{x}_0$ such that $\mathbf{z}=d^{-1}(\mu-\mathbf{x}_0)$ is the north pole and express the integral of $G(\mathbf{x})$ over $\mathbb{R}^3$ with respect to it.
Since
\begin{equation*}
	\|\mathbf{x}-\mu\|^2=\|(\mathbf{x}-\mathbf{x}_0)-(\mu-\mathbf{x}_0)\|^2=r^2-2dr\cos\theta+d^2=(r-d\cos\theta)^2+d^2\sin^2\theta,
\end{equation*}
we have
\begin{align*}
	\int_{\mathbb{R}^3} G(\mathbf{x})\,d\mathbf{x} &= \int_0^{2\pi}d\phi\int_0^{\pi}d\theta \int_0^{\infty} \exp\!\left(-\frac{(r-d\cos\theta)^2+d^2\sin^2\theta}{2\sigma^2}\right)r^2\sin\theta\,dr \\
	&=  \int_{\Omega}\int_0^{\infty} \exp\!\left(-\frac{(r-d\cos\theta)^2+d^2\sin^2\theta}{2\sigma^2}\right)r^2 \,drd\omega,
\end{align*}
where $\Omega$ denotes the unit sphere, and $\cos\theta=\omega\cdot \mathbf{z}$.
\\

We now compute
\begin{equation*}
	f_o(\omega)=\int_0^{\infty} \exp\!\left(-\frac{(r-d\cos\theta)^2+d^2\sin^2\theta}{2\sigma^2}\right)r^2 \,dr.
\end{equation*}
Note that
\begin{equation*}
	r^2=(r-d\cos\theta)^2+2d\cos\theta(r-d\cos\theta)+d^2\cos^2\theta.
\end{equation*}
By the change of variable $s=r-d\cos\theta$, we have
\begin{equation*}
	f_o(\omega)=\int_{-d\cos\theta}^{\infty} \exp\!\left(-\frac{s^2+d^2\sin^2\theta}{2\sigma^2}\right)\!(s^2+2d\cos\theta\cdot s+d^2\cos^2\theta) \,ds.
\end{equation*}
By integrating by parts using
\begin{equation*}
	\int \exp\!\left(-\frac{s^2+d^2\sin^2\theta}{2\sigma^2}\right)\!s\,ds=-\sigma^2\exp\!\left(-\frac{s^2+d^2\sin^2\theta}{2\sigma^2}\right)+C,
\end{equation*}
we have
\begin{align*}
 \int_{-d\cos\theta}^{\infty} \exp\!\left(-\frac{s^2+d^2\sin^2\theta}{2\sigma^2}\right)\!(s^2+2d\cos\theta\cdot s) \,ds 
	=& \left[-\sigma^2 \exp\!\left(-\frac{s^2+d^2\sin^2\theta}{2\sigma^2}\right)\!(s+2d\cos\theta)\right]_{-d\cos\theta}^{\infty} \\
	& +\sigma^2 \int_{-d\cos\theta}^{\infty} \exp\!\left(-\frac{s^2+d^2\sin^2\theta}{2\sigma^2}\right) ds \\
	=& \ \sigma^2 e^{-\frac{d^2}{2\sigma^2}} d\cos\theta +\sigma^2 \int_{-d\cos\theta}^{\infty} \exp\!\left(-\frac{s^2+d^2\sin^2\theta}{2\sigma^2}\right) ds.
\end{align*}
On the other hand, we have
\begin{align*}
	\int_{-d\cos\theta}^{\infty} \exp\!\left(-\frac{s^2+d^2\sin^2\theta}{2\sigma^2}\right) ds &= e^{-\frac{d^2\sin^2\theta}{2\sigma^2}} \left\{\int_0^{\infty}e^{-\frac{s^2}{2\sigma^2}}\,ds+\int_0^{d\cos\theta}\!\!e^{-\frac{s^2}{2\sigma^2}}\,ds\right\} \\
	&= \sqrt{2}\,\sigma \,e^{-\frac{d^2\sin^2\theta}{2\sigma^2}} \left\{\int_0^{\infty}e^{-t^2}\,dt+\int_0^{\frac{d\cos\theta}{\sqrt{2}\,\sigma}}\!\!e^{-t^2}\,dt\right\} \\
	&= \sqrt{\frac{\pi}{2}}\,\sigma\, e^{-\frac{d^2\sin^2\theta}{2\sigma^2}} \!\left(1+\operatorname{erf}\!\left(\frac{d\cos\theta}{\sqrt{2}\sigma}\right)\!\right),
\end{align*}
where we have applied the change of variable $t=s/(\sqrt{2}\,\sigma)$.
Putting these together, we obtain
\begin{align*}
	f_o(\omega) &= \sigma^2 e^{-\frac{d^2}{2\sigma^2}} d\cos\theta +(\sigma^2+d^2\cos^2\theta) \int_{-d\cos\theta}^{\infty} \exp\!\left(-\frac{s^2+d^2\sin^2\theta}{2\sigma^2}\right) ds \\
	&= \sigma^2 e^{-\frac{d^2}{2\sigma^2}} d\cos\theta +\sqrt{\frac{\pi}{2}}\,\sigma\, e^{-\frac{d^2\sin^2\theta}{2\sigma^2}}\,(\sigma^2+d^2\cos^2\theta) \!\left(1+\operatorname{erf}\!\left(\frac{d\cos\theta}{\sqrt{2}\sigma}\right)\!\right),
\end{align*}
as desired.


\section{Proof of sampling algorithm}

Now, define $\Phi_{\mathbf{x}_0}:\mathbb{R}^3\setminus\{\mathbf{x}_0\}\rightarrow \Omega$ by
\begin{equation*}
	\Phi_{\mathbf{x}_0}(\mathbf{x})=\frac{1}{\|\mathbf{x}-\mathbf{x}_0\|}(\mathbf{x}-\mathbf{x}_0) \qquad (\mathbf{x}\in\mathbb{R}^3\setminus\{\mathbf{x}_0\}).
\end{equation*}
Let us sample $\mathbf{x}$ following \eqref{3DGaussian}.
Thus, for every random variable $X$ on $\mathbb{R}^3$,  we have
\begin{equation*}
	\mathbb{E}[X]=\frac{1}{\sqrt{(2\pi\sigma^2)^3}}\int_{\mathbb{R}^3} X(\mathbf{x})G(\mathbf{x})\,d\mathbf{x}.
\end{equation*}
We discard $\mathbf{x}$ if we happen to get $\mathbf{x}=\mathbf{x}_0$.
Then, for every random variable $X'$ on $\Omega$, we have
\begin{align*}
	\mathbb{E}[X'\circ \Phi] &= \frac{1}{\sqrt{(2\pi\sigma^2)^3}}\int_{\mathbb{R}^3} X'(\Phi(\mathbf{x}))G(\mathbf{x})\,d\mathbf{x} \\
	&= \frac{1}{\sqrt{(2\pi\sigma^2)^3}}\int_{\Omega} \int_0^{\infty} X'(\omega)\, \exp\!\left(-\frac{(r-d\cos\theta)^2+d^2\sin^2\theta}{2\sigma^2}\right)r^2 \,drd\omega \\
	&= \frac{1}{\sqrt{(2\pi\sigma^2)^3}}\int_{\Omega} X'(\omega) \int_0^{\infty} \exp\!\left(-\frac{(r-d\cos\theta)^2+d^2\sin^2\theta}{2\sigma^2}\right)r^2 \,drd\omega \\
	&= \frac{1}{\sqrt{(2\pi\sigma^2)^3}}\int_{\Omega} X'(\omega) f_o(\omega)\, d\omega,
\end{align*}
where we applied the change of variables $\mathbf{x}-\mathbf{x}_0=r\omega$.
Note that $\omega=\Phi(\mathbf{x})$ (when $r>0$).
Hence, it follows that $\Phi(\mathbf{x})$ obeys the spherical distribution $f_o$ (after normalization).

\end{document}